\documentclass[12pt]{iopart}
\usepackage{color}
\usepackage{graphics}
\usepackage{epsf}
\usepackage{graphicx,epsfig}
\usepackage{amssymb}
\def\dh{\dot{H}}
\def\doo{\dot{\phi}}
\def\ddo{\ddot{\phi}}
\def\cs2{c_{s}^{2}}

 \def\al{\alpha}
 \def\b{\beta}
 \def\ga{\gamma}
 \def\de{\delta}
 \def\De{\Delta}
 \def\ep{\varepsilon}
 \def\ze{\zeta}
 \def\th{\theta}

 \def\ph{\varphi}

 \def\df{\delta\phi}
 \def\p{\partial}
 \def\ddf{\delta\dot{\phi}}
 \def\alo{\alpha_{1}}
 \def\alt{\alpha_{2}}
 \def\tho{\theta_{1}}
 \def\tht{\theta_{2}}
 \def\bt{\bigtriangledown}
 \def\dP{\dot{\phi}}
 \def\epr{\eta^{'}}
 \def\eps{\eta^{''}}

\def\f{\phi}
\def\fd{\dot{\f}}
 \def\be   {\begin{equation}}   \def\ee   {\end{equation}}
 \def\ba   {\begin{array}}      \def\ea   {\end{array}}
 \def\bea  {\begin{eqnarray}}   \def\eea  {\end{eqnarray}}
 \def\bean {\begin{eqnarray*}}  \def\eean {\end{eqnarray*}}
\begin{document}

\title{One-loop corrections to the power spectrum in general single-field inflation}

\vspace{0.8cm}

\author{Nicola Bartolo$^{1,2}$, Emanuela Dimastrogiovanni$^{1,2}$ and Alberto Vallinotto$^{3}$}
\vspace{0.4cm}
\address{$^1$ Dipartimento di Fisica ``G. Galilei'', Universit\`{a} degli Studi di 
Padova, \\ via Marzolo 8, I-35131 Padova, Italy} 
\address{$^2$ INFN, Sezione di Padova, via Marzolo 8, I-35131 Padova, Italy}
\address{$^3$ Center for Particle Astrophysics, Fermi National Accelerator Laboratory, P.O. Box 500, Kirk Rd. \& Pine St., Batavia, IL 60510-0500 USA}
\eads{\mailto{nicola.bartolo@pd.infn.it}, \mailto{dimastro@pd.infn.it} and \mailto{avalli@fnal.gov}}

\date{\today}
\vspace{1cm}

\begin{abstract}
We perfom a thorough computation of the one-loop corrections from both scalar and tensor degrees of freedom to the power spectrum of curvature fluctuations for 
non-canonical Lagrangians in single-field inflation. We consider models characterized by a small sound speed $c_{s}$, which produce large non-Gaussianities. 
As expected, the corrections turn out to be inversely proportional to powers of $c_{s}$; evaluating their amplitudes it is then possible to derive some theoretical 
bounds on the sound speed by requesting the conditions necessary for perturbation theory to hold.

\end{abstract}

\maketitle

\section{Introduction}

The Cosmic Microwave Background (CMB) represents one of the most important sources of information for cosmology. CMB fluctuations from the last scattering surface contain precious indications about early Universe physics. The Wilkinson Microwave Anisotropy Probe (WMAP) and the Planck satellite have been helping collect all these ``signals'' from the ealy Universe and with increasing accuracy and sensitivity \cite{smoot92,bennett96,gorski96,wmap3,wmap5,SSZ}. Observations have confirmed that many of the existing models share a satisfactory agreement with experiments. The new data are expected to further help understanding what is the correct model. Great expectations rely on non-Gaussianity \cite{review,prop} and on the possibility to improve the measurements of the two-point function of cosmic fluctuations. This has encouraged the theoretical work on three- and four-point correlation functions of the cosmological fluctuations and on the study of loop corrections to the tree-level correlators as well. Both the former and the latter are in fact deeply model-dependent since they are related to the specific interactions characterizing the primordial fields. \\

Loop corrections have received a renewed interest since the works~\cite{weinberg1,weinberg2}, where one of the main issues to be addressed is whether loop corrections can carry information about the history of 
inflation and whether they can become large. Among the various results is that the power spectrum of the curvature perturbations acquires a new (compared to the tree-level) 
dependence from the external momentum, i.e. of a logarithmic kind ($\simeq lnk$), arising from ultraviolet divergences (see also \cite{sloth,sloth2,seery1,seery2,Dimastrogiovanni:2008af,Giddings:2010nc,Giddings:2010ui}). In fact, recently 
the authors of Ref.~\cite{SZnew} showed that such a dependence must be computed with care, considering a cutoff of the order $a(t_k) \Lambda$, where $t_k$ is the time of horizon exit for the mode $k$ during inflation, 
 so that the logarithmic dependence $ln(k/a(t_k) \Lambda)$ turns out to be actually fixed by $ln(H/\Lambda)$ \footnote{Notice that the results of \cite{sloth,sloth2,seery1,seery2,Dimastrogiovanni:2008af,Giddings:2010nc,Giddings:2010ui} have been obtained using different regularization techniques. In particular, in \cite{Dimastrogiovanni:2008af} we employed a cutoff on comoving momenta, rather than resorting to dimensional regularization.}. However, loop corrections to the cosmological correlators have attracted a lot of interest mostly because of the following reason: they are characterized by infrared divergences ($\simeq lnkL$, where $L^{-1}$ is an infrared cutoff), and by the presence of secular terms growing like the logarithm of the scale factor ($\simeq ln(-k\eta)$), which diverge in the limit of late times. A debate has then been stimulated about the significance of these divergences. \\
As to the $lnkL$ terms, an interpretation for them has been proposed in terms of the so called ``small'' and ``large'' boxes: the former corresponds to considering an (inverse) infrared cutoff $L$ of the size of our observable Universe, which also implies that $k$ is not much smaller than $L^{-1}$; on the other hand, a ``large'' box corresponds to choosing a cutoff $L\gg k^{-1}$. When dealing with observable quantities, the natural cutoff should be the ``small'' box, whereas any correlator computed considering the ``large'' is expected to be interpreted as some quantity that is unaccessible by measurements \cite{box4,box5} (see also~\cite{Enqvistetal}). In this view, the quantities $\simeq lnkL$ are not large and therefore should not be interpreted as actual divergences. Another approach might consist in a resummation of these divergences through Renormalization Group techniques (see the interesting example of Ref.~\cite{TS}). \\
As to the $ln(-k\eta)$ terms, their presence was also discussed in \cite{weinberg1,weinberg2}, where it was pointed out that these terms are generally expected in the loop corrections, unlike any positive power law dependence from the scale factor. This divergence appears to be fictional since observations are ``made at a finite time'' \cite{seery2}, i.e. $\eta\not= 0$. Regarding this issue, the $\delta$N formalism \cite{deltaN1,deltaN2,deltaN3} offers a nice way out because two different times are involved: an initial one, at which the correlators of the scalar field is evaluated, an a final one, which corresponds to the time of observations. The final time is a late time, the initial can be arbitrarily chosen and a very convenient choice is to set it just a few e-foldings after horizon crossing. This means that the quantity $|k\eta|$ is expected to be quite close to one. \\
When it comes to computing resummations of loop corrections, an interesting discussion is provided by \cite{Burgess:2009bs}, where a ``Dynamical Renormalization Group'' (DRG) technique is developed in order to support the perturbative approach at late times (this technique is also employed in order to improve the IR behaviour by resumming late time correlators). See also \cite{Seery:2010kh} for a recent review on inflationary loop correction. \\

Loop correction have been computed for the power spectrum of curvature fluctuations in the basic single-field slow-roll model: in \cite{seery1,seery2}, these corrections were computed considering the fluctuations of the background metric, setting however the tensor modes to zero for simplicity; in \cite{Dimastrogiovanni:2008af} the tensor modes were fully accounted for and a complete computation was carried out in order to evaluate the nature of the tensor mode loop corrections, which were proved to be characterized by infrared logarithmic divergences and to be of the same order of magnitude as the scalar corrections. This showed that a full computation of one-loop corrections in standard single-field slow-roll inflation requires the knowledge of the tensor contributions \footnote{Notice that in Eq.~(62) of \cite{Dimastrogiovanni:2008af} the numerical coefficient multiplying the $\ln kL$ term has been corrected in \cite{Seery:2010kh}. In particular $f_{2}$ is equal to a factor 2. We thank David Seery for pointing this out.}. Loop corrections have also been computed in more general theories of inflation such as canonical multifield \cite{Adshead:2008gk} and multiple Dirac-Born-Infeld models (DBI) \cite{Gao:2009fx} (where a multifield DBI model was considered and the loop corrections arising from cross-correlations between adiabatic and non-adiabatic modes were computed for the scalar fields power spectrum). The latter models are particularly interesting because they are part of a larger class of models characterized by 
an inflaton lagrangian that is a generic function $P(X,\phi)$ of the inflaton field $\phi$ and its first derivatives, $X\equiv-\frac{1}{2}g^{\mu\nu}\p_{\mu}\phi\p_{\nu}\phi$, 
with a sound speed $c_{s}$ 
which can be small. In these models the interaction Hamiltonian is generally characterized by terms that are proportional to inverse powers of $c_{s}$ and that can be therefore responsible for non-negligible amplitude of the three- and four-point functions \cite{DBIsky,Chen:2006nt,Huang:2006eha,Chen:2009bc,Arroja:2008ga,Arroja:2008yy,Arroja:2009pd,Mizuno:2009mv}, as well as loop corrections to the two-point function. For models of inflation that comprise higher derivatives of the inflaton field in the Lagrangian see also \cite{ArkaniHamed:2003uz,Cheung,SSZ,Bartolo:2010bj}. For more works on inflationary loop corrections see also \cite{Boyanovsky:2005px,Boyanovsky:2005sh,vanderMeulen:2007ah,Byrnes:2007tm,Adshead:2008gk,Burgess:2010dd,SZnew}. \\

In this paper we are going to study single-field $P(X,\phi)$ models of inflation and formulate a complete computation of the loop corrections to the power spectrum of the curvature perturbations, both in the spirit of \cite{Dimastrogiovanni:2008af}, i.e. by considering all of the metric fluctuations (including tensor modes), and in the spirit of \cite{Leblond:2008gg}, i.e. using the loop computation and the perturbative expansion conditions in an attempt to extract some constraints for the parameters of the theory and for the sound speed in particular.\\

The paper is organized as follows: in Sec.~\ref{beom} we list the background equations of motion and derive some fundamental expressions for the parameters of the theory (summarized in Table 1); in Sec.~\ref{delta n} we recall the $\delta$N formalism, with some comments on its applicability to generic $P(X,\phi)$ inflation models, and we employ it for the calculation of the power spectrum of $\zeta$; in Sec.~\ref{perturbation-L} we review the main steps for deriving the interaction Hamiltonian and report the leading order terms that will enter the loop calculations; in Sec.~\ref{actualcomp} we report the order of magnitude of the scalar and of the tensor loop diagrams; in Sec.~\ref{finalresults} we present our final results; finally, in Sec.~7 we draw our conclusions. The calculations we performed are quite lengthy, so many of the details are left to the Appendices: in Appendix A we solve the Hamiltonian and momentum constraints for $P(X,\phi)$ models; in Appendix B we present the complete expressions for the third and fourth order Lagrangians, including tensor modes; in Appendices C and D we review the computation of, respectively, two-vertex and one-vertex loop diagrams; finally in Appendix E we provide some of the polarization tensor equations that are needed for tensor loop calculations.

%%%%%%%%%%%%%%%%%%%%%%%%%%%%%%%%%%%%%%%%%%%%%%%%%%%%%%%%%%%%%%%%%%%%%%%%%%%%%%%%%%%%%%%%%%%%%%%%%%%%%%%%%%%%%%%%%%%%%
%%%%%%%%%%%%%%%%%%%%%%%%%%%%%%%%%%%%%%%%%%%%%%%%%%%%%%%%%%%%%%%%%%%%%%%%%%%%%%%%%%%%%%%%%%%%%%%%%%%%%%%%%%%%%%%%%%%%%
%%%%%%%%%%%%%%%%%%%%%%%%%%%%%%%%%%%%%%%%%%%%%%%%%%%%%%%%%%%%%%%%%%%%%%%%%%%%%%%%%%%%%%%%%%%%%%%%%%%%%%%%%%%%%%%%%%%%%
%%%%%%%%%%%%%%%%%%%%%%%%%%%%%%%%%%%%%%%%%%%%%%%%%%%%%%%%%%%%%%%%%%%%%%%%%%%%%%%%%%%%%%%%%%%%%%%%%%%%%%%%%%%%%%%%%%%%%
%%%%%%%%%%%%%%%%%%%%%%%%%%%%%%%%%%%%%%%%%%%%%%%%%%%%%%%%%%%%%%%%%%%%%%%%%%%%%%%%%%%%%%%%%%%%%%%%%%%%%%%%%%%%%%%%%%%%%
%%%%%%%%%%%%%%%%%%%%%%%%%%%%%%%%%%%%%%%%%%%%%%%%%%%%%%%%%%%%%%%%%%%%%%%%%%%%%%%%%%%%%%%%%%%%%%%%%%%%%%%%%%%%%%%%%%%%%
%%%%%%%%%%%%%%%%%%%%%%%%%%%%%%%%%%%%%%%%%%%%%%%%%%%%%%%%%%%%%%%%%%%%%%%%%%%%%%%%%%%%%%%%%%%%%%%%%%%%%%%%%%%%%%%%%%%%%
%%%%%%%%%%%%%%%%%%%%%%%%%%%%%%%%%%%%%%%%%%%%%%%%%%%%%%%%%%%%%%%%%%%%%%%%%%%%%%%%%%%%%%%%%%%%%%%%%%%%%%%%%%%%%%%%%%%%%
%%%%%%%%%%%%%%%%%%%%%%%%%%%%%%%%%%%%%%%%%%%%%%%%%%%%%%%%%%%%%%%%%%%%%%%%%%%%%%%%%%%%%%%%%%%%%%%%%%%%%%%%%%%%%%%%%%%%%

\section{Background equations of motion}\label{beom}

We will consider a class of theories described the following action
\bea\label{beginning}
S=\frac{1}{2}\int d^4x \sqrt{-g}\left[m_{P}^{2}R+2P(X,\phi)\right]
\eea
where $R$ is the four dimensional Ricci scalar, $m_{P}\equiv(8 \pi G)^{-1/2}$ and $P$ is a generic function of the inflaton field $\phi$ and its first derivatives 
$X\equiv-\frac{1}{2}g^{\mu\nu}\p_{\mu}\phi\p_{\nu}\phi$. \\
The background equations are \cite{Chen:2006nt,Seery:2005wm}
\bea\label{a}
2\dh+3H^2=-P ,\\
H^2=\frac{1}{3}\left[2XP_{X}-P\right],\label{b}\\
\dot{X}\left(P_{X}+2XP_{XX}\right)+2\sqrt{3}{\left(2XP_{X}-P\right)}^{1/2}XP_{X}=\sqrt{2X}\left(P_{\phi}-2XP_{X\phi}\right),\label{c}
\eea
where a dot indicates a derivative w.r.t. cosmic time, $P_{\phi}=\partial P/\partial \phi$ and we define $X \equiv {\dot{\phi}}^{2}/2$ 
(we have set $m_{P}=1$). In the case of standard single-field models of slow-roll inflation $P=X-V(\phi)$ where 
$V(\phi)$ is the potential for the scalar field.\\

Let us introduce some ``slow-variation'' parameters
\bea
\ep\equiv -\frac{\dot{H}}{H^2}\label{d},\quad\quad\quad\quad\quad\quad \eta \equiv \frac{\dot{\ep}}{\ep H}\label{e}\, ,
\eea
which are assumed to be smaller than one, as for the slow-roll parameters in the standard case. However, in $P(X,\phi)$ theories it is not generally correct to talk about ``slow-roll''; in fact, $P$ is only known to be a generic function of $X$ and $\phi$, therefore the smallness of $\ep$ and $\eta$ does not necessarily indicate that ${\doo}^{2} \ll H^{2}$ and $|\ddo| \ll |3 H  \doo|$.   \\
\noindent The speed of sound in these models has the following expression
\bea\label{sp}
\cs2=\frac{P_{X}}{P_{X}+2XP_{XX}}\label{ii}.
\eea
We do not assign a specific value to $c_{s}$ at this stage, the only assumption we make is that its derivative is small, more precisely
\bea
s \equiv \frac{\dot{c_{s}}}{c_{s} H}\label{i}\ll 1,
\eea
which appears as a natural choice dictated by the expression of the spectral index of scalar perturbations for these models 
($n_s-1=-2\epsilon-\eta-s$, see, e.g., Ref.~\cite{Garriga:1999vw,Chen:2006nt}). 
The three parameters $\ep$, $\eta$ and $s$ in $P(X,\phi)$ theories are dubbed as ``slow variation'' parameters because $\ep$ and $\eta$ indicate that the variation of $H$ w.r.t. time is quite slow, i.e. the expansion is quasi de-Sitter; similarly, a small $s$ indicates a slowly varying sound speed.\\

\subsection{Useful relations in terms of slow-variation parameters}
\noindent 
A computation of the inflaton power spectrum up to one-loop requires to expand the Lagrangian up to fourth-order in the perturbations. 
The perturbative expansion of the Lagrangian is expected to be structured as the sum of a very large number of terms, 
each one weighted by coefficients given by the zeroth order part of $P$ and its derivatives w.r.t $X$ and $\phi$. It is then useful to find out the 
dependence of these coefficients from the ``slow-variation'' parameters and from the sound speed $c_s$. The main results of this section are summarized in 
Table~1 where we give the order in slow-variation parameters (and in the sound speed) of the various coefficients.\\

\noindent To begin with, the expression of $XP_{X}$ easily follows from the combination of (\ref{a}) and (\ref{b})
\be\label{ww}
XP_{X}=-\dh,
\ee
which can be plugged in (\ref{d}) to have
\be\label{res2}
XP_{X}=\ep H^2
\ee
(as a check, in the standard case $P_{X}=1$ and the equation above gives $\ep=({\doo}^{2})/(2 H^2)$). \\
The expression of $X^2P_{XX}$ in terms of slow variation parameters and of the sound speed is derived just as easily from the combination of Eqs.~(\ref{sp}) and (\ref{res2}), i.e.
\bea\label{www}
X^2P_{XX}=\frac{\ep}{2}\left(\frac{1}{c_{s}^{2}}-1\right)H^{2}.
\eea
The expressions for the higher derivatives of $P$ can be obtained by successive derivations of (\ref{res2}) and (\ref{www}) w.r.t $X$ and $\phi$. We thus expect a dependence not only from the parameters $\ep$, $\eta$, $s$ and $c_{s}$, but also from their derivatives. \\
We will now derive some equations relating the derivatives of $P$ with each other and with $c_{s}$ and the slow-variation parameters. It is useful to define the following combinations of derivatives \cite{Chen:2006nt}
\bea
\Sigma \equiv XP_{X}+2X^2P_{XX} \label{m},\\ 
\lambda \equiv  X^{2}P_{XX}+\frac{2}{3}X^{3}P_{XXX}\label{n},\\
\Pi \equiv X^{3}P_{XXX}+\frac{2}{5}X^4P_{XXXX}\label{o}.
\eea
\noindent From (\ref{res2}) and (\ref{www}), we get 
\bea
\Sigma=\frac{\ep}{c_{s}^{2}}H^{2}.
\eea
There is no such simple expression for $\lambda$ and $\Pi$ in the general case. The former, for instance, can be written as
\be\label{L1}
\lambda=\frac{H^{2} \ep}{c_{s}^{2}}\Big[-\frac{2}{3}\frac{Xc_{sX}}{c_{s}}+\frac{(1-c_{s}^{2})}{6c_{s}^{2}}\Big].
\ee

where we introduced the partial derivatives of the sound speed w.r.t to $X$ and $\phi$ as $\dot{c}_{s}=\dot{X}c_{s X}+\dot{\phi}c_{s\phi}$. 
In order to derive Eq.(\ref{L1}), one has to first take the derivative of Eq.~(\ref{ii}) w.r.t. $X$, 
then multiply both sides by $X$ and use the definition of $\lambda$ in (\ref{n}) and of the sound speed.\\

It will also be convenient for future calculations to decompose $\ep$ as the sum (here 
$\ep_{X}$ is not the partial derivative of $\ep$ w.r.t. $X$ but merely the definition provided in (\ref{clarify}) 
\be\label{f}
\ep=\ep_{\phi}+\ep_{X},
\ee
where
\bea\label{clarify}
\ep_{\phi} \equiv -\frac{\doo}{H^2}\frac{\p H}{\p \phi}\label{g},\quad\quad\quad\quad\quad\quad\quad\quad
\ep_{X} \equiv -\frac{\dot{X}}{H^2}\frac{\p H}{\p X}\label{h}.
\eea
In standard single-field models of inflation $\Sigma$, $\lambda$ and $\Pi$ are equal to zero, whereas $\ep_{\phi}$ and $\ep_{X}$ become

%(\ref{b}) and (\ref{c}) become
%\bea
%3H^2=\frac{{\phi}^{2}}{2}+V,\\
%\ddo+3H\doo+V_{\phi}=0,
%\eea
%so
%\be
%\frac{\p H}{\p \phi}=\frac{V_{\phi}}{6H}
%\ee
%and therefore
\be\label{tt}
\ep_{\phi}=-\frac{\doo}{H^2}\frac{V_{\phi}}{6H}=\frac{\doo}{H^2}\frac{1}{6H}\left(\ddo+3H\doo\right)=\frac{{\doo}^{2}}{2H^2}+\frac{\doo\ddo}{6H^3}
\ee
where $V_{\phi}\equiv \p V/\p \phi$ and
%Deriving (\ref{b}) w.r.t. $X$ we get
%\be
%6H\frac{\p H}{\p X}=P_{X}+2XP_{XX}
%\ee
%which in the standard case reduces to
%\be
%6H\frac{\p H}{\p X}=\frac{1}{6H}
%\ee
%whereas $\dot{X}=\doo\ddo$, so we have
\be\label{ttt}
\ep_{X}=-\frac{1}{H^{2}}\doo\ddo\frac{1}{6H}=-\frac{\doo\ddo}{6H^{3}}.
\ee 
Summing up $\ep_{\phi}$ and $\ep_{X}$ from (\ref{tt}) and (\ref{ttt}), we recover $\ep=({\doo}^{2}/2)(1/H^2)$. If the slow roll condition $|\ddo| \ll |3H\doo|$ is applied, $\ep_{X}$ is evidently subdominant w.r.t. $\ep_{\phi}$. \\
For a general $P(X,\phi)$ theory one finds
\bea\label{ssss}
\ep_{X}=\frac{\ep}{3(1+c_{s}^{2})}\left[2\ep-\eta+\frac{\dot{\phi} X P_{X\phi}}{\ep H^3}\right].
\eea
Equivalently we have \footnote{E.\,D. thanks Sarah Shandera for corresponding about the contents of this section and for spotting an error in the first version of these calculations.}
\bea\label{sssss}
\ep_{X}=\frac{\ep}{3(1+c_{s}^{2}-4Xc_{s}c_{sX})}\left[2(\ep+s)-\eta+\frac{\dot{\phi} X P_{X\phi}}{\ep H^3}-2\frac{\dot{\phi}c_{s\phi}}{Hc_{s}}\right].
\eea
Eq.~(\ref{ssss}) can be derived by using Eqs.~(\ref{b}), (\ref{c}), (\ref{g}) and noticing that (\ref{b}) implies
\be
P_{\phi}-2XP_{X\phi}=-6H\frac{\p H}{\p \phi},
\ee
so that
\be\label{res}
\dot{X}=-6 H \cs2 X \frac{\ep_{X}}{\ep} \,.
\ee
The previous equation can also be written as
\bea\label{Res}
\dot{X}=\frac{2XH c_{s}^{2}}{1+c_{s}^{2}}\left[\eta-2\ep-\frac{\dot{\phi}P_{X\phi}}{HP_{X}}\right]\, ,
\eea
by using $\dot{X}=\dot{\phi}\ddot{\phi}$ and 
\bea
\ddot{\phi}=\frac{d}{dt}\left(H\sqrt{\frac{2\ep}{P_{X}}}\right)=\dot{\phi}\left(\frac{\eta H}{2}-\ep H-\frac{\dot{\phi}P_{X\phi}+\dot{X}P_{XX}}{2P_{X}}\right).
\eea
Eq.~(\ref{ssss}) follows from equating (\ref{res}) and (\ref{Res}). In order to derive Eq.~(\ref{sssss}), we simply take the time derivative of $\Sigma$ in Eq.~(\ref{m}), i.e.
\bea
\dot{\Sigma}=\dot{X}\Sigma_{X}+\dot{\phi}\Sigma_{\phi}
\eea
and we use the definition of the sound speed, together with Eq.~(\ref{res}). It is easy to verify that (\ref{ssss}) and (\ref{sssss}) agree with each other.\\
Other useful and easily verifiable equations are the following
\bea\label{iconto}
\frac{\dot{\phi}P_{\phi}}{2H^3}=3\ep-2\ep^2+\ep\eta+3\ep_{X}c_{s}^{2} \\\label{riconto}
\frac{\dot{\phi}X P_{X\phi}}{H^3}= \ep\eta-2\ep^2+3\ep_{X}(1+c_{s}^{2})
\eea
\bea
\frac{\dot{\phi}X^2 P_{XX\phi}}{H^3}&=&\frac{\ep}{c_{s}^{2}}\Bigg[-s+\frac{3}{2}\frac{\ep_{X}}{\ep}(c_{s}^{2}-1)+\frac{9\lambda}{H^2}c_{s}^{4}\frac{\ep_{X}}{\ep^2}\nonumber\\&+&\frac{1}{2}(c_{s}^{2}-1)\left(\eta-2\epsilon+3\frac{\ep_{X}}{\ep}(1+c_{s}^{2})\right)\Bigg] 
\eea

\subsubsection{ Some worked examples}
\label{examples}
%\vspace
\smallskip $\quad\quad\quad\quad\quad\quad\quad\quad\quad\quad\quad\quad\quad\quad\quad\quad\quad\quad\quad\quad\quad\quad\quad\quad\quad\quad\quad\quad\quad\quad\quad\quad\quad\quad\quad\quad\quad\quad\quad\quad\quad\quad\quad\quad\quad\quad\quad\quad\quad\quad\quad\quad\quad\quad\quad\quad\quad\quad$\\
\noindent In order to get a hint of the typical values for these quantities in some of the known models, we will derive the values of $\ep_{X}$ and $c_{sX}$ 
for three cases: 1) DBI model; 2) $P(X,\phi)=f(X)+g(\phi)$ models (where $f$ and $g$ are generic functions of their arguments), 
of which the canonical slow-roll Lagrangian represents a special case; 3) $P(X,\phi)=f(X)g(\phi)$ models (of which K-inflation is a special case).

\begin{itemize}

\item DBI model.

\bea
P_{DBI}(X,\phi)=-\frac{\sqrt{1-2Xf(\phi)}}{f(\phi)}+\frac{1}{f(\phi)}-V(\phi).
\eea
In this model we have 
\bea\label{css}
c^{2}_{s}=1-2Xf
\eea
and 
\bea\label{px}
P_{X}=\frac{1}{c_{s}},
\eea
so
\bea\label{alternativa}
\ep=\frac{X}{c_{s}H^2}\, ,
\eea
and 
\bea\label{fff}
f=\frac{1-c_{s}^{2}}{2X}=\frac{1-c_{s}^{2}}{2\ep c_{s}H^{2}}.
\eea
Let us compute the order of $\ep_{X}$. From Eq.~(\ref{ssss}), we know that this can be derived by computing $\dot{\phi} X P_{X\phi}$. Taking the derivative of $P_{X}$ in Eq.~(\ref{px}) w.r.t. the scalar field we get
\bea\label{dec}
\dot{\phi} X P_{X\phi}=\frac{\dot{\phi}f_{\phi}X^2}{(1-2Xf)^{3/2}}=\frac{\dot{f}X^2}{c_{s}^{3}}.
\eea
Replacing the derivative of Eq.~(\ref{fff}) w.r.t. time 
\bea
\dot{f}=c_{s}\left(\frac{\eta-s-2\ep}{2\ep H}\right)+\frac{1}{c_{s}}\left(\frac{2\ep-s-\eta}{2\ep H}\right),
\eea
Eq.~(\ref{dec}) becomes
\bea
\dot{\phi} X P_{X\phi}=\dot{f}\frac{\ep^2 H^{4}}{c_{s}}\simeq \frac{\ep^2}{c_{s}^2}H^3
\eea
where, in the last line, we assumed that $\eta\simeq\mathcal{O}(\ep)\simeq\mathcal{O}(s)$ and that $c_{s}<1$ (we are interested in small sound speed values). 
Replacing this result in (\ref{ssss}) we get $\ep_{X}\simeq\mathcal{O}(\ep^2/c_{s}^2)$. As a double-check, we can calculate the expression of $\ep_{X}$ by taking the time derivative of (\ref{alternativa}) and using the definition of $\eta$; the result is
\bea
\ep_{X}=-\frac{\ep}{6c_{s}^{2}}\left(\eta+s-2\ep\right).
\eea
Let us now evaluate $c_{sX}$. Taking the derivative of (\ref{css}) w.r.t. $X$ we get
\bea
2c_{s}c_{sX}=-2f=\frac{c_{s}^{2}-1}{X}\, ,
\eea
then $Xc_{s}c_{sX}\simeq\mathcal{O}(1)$ if $c_{s}<1$. \\

\item $P(X,\phi)=f(X)+g(\phi)$ models.\\  
In these models $P_{X \phi}=0$, therefore $\ep_{X}\simeq\mathcal{O}(\ep^2)$ from Eq.~(\ref{ssss}). The sound speed is independent of $\phi$
\bea
c_{s}^{2}=\frac{f_{X}}{f_{X}+2Xf_{XX}},
\eea
so from $\dot{c}_{s}=\dot{X}c_{sX}$ and using Eq.~(\ref{ssss}) and (\ref{res}) we get
\bea\label{soloprima}
2Xc_{s}c_{sX}=\left(1+c_{s}^{2}\right)\left(\frac{s}{\eta-2\ep}\right)\, .
\eea
This brings $Xc_{s}c_{sX}\simeq\mathcal{O}(1)$ if $s\sim\mathcal{O}(\ep)$.\\ 

\item $P(X,\phi)=f(X)g(\phi)$ models.\\ 
The order of $\ep_{X}$ can be computed knowing that 
\bea
\dot{\phi}XP_{X\phi}=\ep H^2\frac{\dot{g}}{g},
\eea
so 
\bea
\ep_{X}=\frac{\ep}{3(1+c_{s}^{2})}\left(2\ep-\eta+\frac{\dot{g}}{gH}\right).
\eea
The first of Eqs.~(\ref{soloprima}) also applies for these models, so $Xc_{s}c_{sX}\simeq\ep s/\ep_{X}$. In a subclass of K-inflation models, known as ``power-law K-inflation'' \cite{ArmendarizPicon:1999rj,Garriga:1999vw} 
\bea\label{ex}
P(X,\phi)=\frac{4}{9}\left(\frac{4-3\gamma}{\gamma^2\phi^2}\right)\left(-X+X^2\right),
\eea
where $\gamma$ is a constant. The sound speed in these models is given by
\bea
c_{s}^{2}=\frac{\gamma}{8-3\gamma},
\eea
so the regime of small $c_{s}$ corresponds to considering small values of $\gamma$. In this regime, it is possible to show that $\dot{g}/gH\simeq \gamma$, from which we have $\ep_{X}\simeq\ep\times\mathcal{O}(\ep+c_{s}^2)\simeq\mathcal{O}(\ep^2)$ (it is easy to verify that $\ep\simeq c_{s}^{2}\simeq\gamma$ for (\ref{ex})). 
\end{itemize}

\noindent We now know what are the typical values of $\ep_{X}$ and $c_{sX}$ in some of the known $P(X,\phi)$ models. Let us see how this information can be used to derive the orders of magnitude of the remaining derivatives of $P$.\\
Let us begin with the derivatives w.r.t. $X$. From the definition of the sound speed we have
\bea\label{qui1}
X^2 P_{XX}=\frac{XP_{X}}{2}\left(\frac{1}{c_{s}^{2}}-1\right)\, ,
\eea
so
\bea\fl
X\frac{d}{dX}\left(X^2 P_{XX}\right)=\frac{1}{2}\left(\frac{1}{c_{s}^{2}}-1\right)\left(XP_{X}+X^2 P_{XX}\right)+\frac{XP_{X}}{2}\left(\frac{-2Xc_{s}c_{sX}}{c_{s}^{4}}\right).
\eea
This can be equated to 
\bea
X\frac{d}{dX}\left(X^2 P_{XX}\right)=2X^2 P_{XX}+X^3 P_{XXX},
\eea
in order to solve for $P_{XXX}$. Using $\ep=XP_{X}/H^2$, we get
\bea\label{qui2}
X^3 P_{XXX}=\frac{\ep H^2}{2}\left[\left(1-\frac{1}{c_{s}^{2}}\right)+\frac{1}{2}\left(1-\frac{1}{c_{s}^{2}}\right)^2-\frac{2Xc_{s}c_{sX}}{c_{s}^{4}}\right].
\eea
In a similar way, we can derive
\bea\fl
X^4 P_{XXXX}=&-&3X^3 P_{XXX}+\frac{\ep H^2}{4}\left(1+\frac{1}{c_{s}^{2}}\right)\left[\left(1-\frac{1}{c_{s}^{2}}\right)+\frac{1}{2}\left(1-\frac{1}{c_{s}^{2}}\right)^2-\frac{2Xc_{s}c_{sX}}{c_{s}^{4}}\right]\nonumber\\\fl
&+&\frac{\ep H^2}{2}\left[\left(1-\frac{1}{c_{s}^{2}}\right)\frac{2Xc_{s}c_{sX}}{c_{s}^{4}}-\frac{2X^2c_{s}c_{sXX}}{c_{s}^{4}}+\frac{6X^2c_{s}^{2}c_{sX}^{2}}{c_{s}^{6}}\right],\eea
where $c_{sXX}\equiv d^2c_{s}/dX^2$. Notice that the equations above are valid for a generic $P(X,\phi)$.\\
The remaining derivatives can be obtained in a similar way. For instance, we know from Eq.~(\ref{iconto}) the expression of $\dot{\phi}P_{\phi}$ as a function of $\ep$, $\eta$ and $\ep_{X}$. By taking  the derivative of (\ref{iconto}) w.r.t. $\phi$ and using (\ref{riconto}), it is possible to easily derive the order of magnitude of $P_{\phi\phi}$ in terms of the flow parameters $\ep$, $\eta$ and $s$, $\epsilon_{X}$ and the first derivatives of $\eta$ and $\ep_{X}$.\\

\begin{table}[t]
\label{Tslow}
\centering
\caption{Order of magnitude of the partial derivatives of $P$ for three classes of $P(X,\phi)$ models. Class I is identified by Eqs.~(\ref{condi2}) and (\ref{incomune}) and includes DBI-like models. 
Class II is identified by Eqs.~(\ref{condi1}) and (\ref{incomune}). Class III corresponds to the model described by Eq.~(\ref{ex}).\\}
\begin{tabular}{|c|c|c|c|}\hline
${}$ & class I & class II & class III \\
\hline\hline
$X^{}P_{X}$ & $(\ep)H^{2}$  & $(\ep)H^{2}$& $(\ga)H^{2}$ \\
\hline
$X^{2}P_{XX}$ & $(\ep/c_{s}^{2})H^{2}$ & $(\ep/c_{s}^{2})H^{2}$ & $H^{2}$ \\
\hline
$X^{3}P_{XXX}$& $(\ep/c_{s}^{4})H^{2}$ & $(\ep/c_{s}^{4})H^{2}$ & $0$\\
\hline
$X^{4}P_{XXXX}$ & $(\ep/c_{s}^{6})H^{2}$  & $(\ep/c_{s}^{6})H^{2}$ & $0$ \\
\hline
$\dot{\phi}^{}P_{\phi}$& $(\ep)H^{3}$ & $(\ep)H^{3}$ & $(\ga)H^{3}$\\
\hline
$\dot{\phi}^{2}P_{\phi\phi}$& $(\ep^{2})H^{4}$ & $(\ep^{2})H^{4}$ & $(\ga^{2})H^{4}$ \\
\hline
$\dot{\phi}^{3}P_{\phi\phi\phi}$& $(\ep^{3})H^{5}$ & $(\ep^{3})H^{5}$ & $(\ga^{3})H^{5}$\\
\hline
$\dot{\phi}^{4}P_{\phi\phi\phi\phi}$ & $(\ep^{4})H^{6}$  & $(\ep^{4})H^{6}$ & $(\ga^{4})H^{6}$  \\
\hline
$\dot{\phi}^{}X^{}P_{X\phi} $& $(\ep^{2}/c_{s}^{2})H^{3}$ & $0$ & $(\ga^{2})H^{3}$  \\
\hline
$\dot{\phi}^{}X^{2}P_{XX\phi} $ & $(\ep^{2}/c_{s}^{4})H^{3}$  & $0$ &$(\ga)H^{3}$ \\
\hline $\dot{\phi}^{}X^{3}P_{XXX\phi}$ & $(\ep^{2}/c_{s}^{6})H^{3}$ & $0$ & $0$\\
\hline
$\dot{\phi}^{2}X^{}P_{X\phi\phi}$ & $(\ep^3/c_{s}^{2})H^{4}$ & $0$ & $(\ga^3)H^{4}$ \\
\hline
$\dot{\phi}^{3}X^{}P_{X\phi\phi\phi}$ & $(\ep^4/c_{s}^{2})H^{5}$ & $0$ & $(\ga^4)H^{5}$ \\
\hline
$\dot{\phi}^{2}X^{2}P_{XX\phi\phi}$ & $(\ep^3/c_{s}^{4})H^{4}$ & $0$ & $(\ga^2)H^{4}$ \\
\hline
\end{tabular}
\label{ordini}
\end{table}

\noindent  For practical reasons, given the large number of terms in the Lagrangian, we will fix the order of magnitude in terms of $\epsilon$, $s$, $\ep_{X}$ and $c_{sX}$. 
In the case where 
\bea\label{condi2}
\ep_{X}\simeq \mathcal{O}\left(\frac{\ep^{2}}{c_{s}^{2}}\right),
\eea
and 
\bea\label{incomune}
Xc_{s}c_{sX}\simeq \mathcal{O}(1),\quad\quad\quad s\simeq \mathcal{O}(\ep),
\eea
we include all DBI models, see first column of Table~\ref{ordini}.\\ 
On the other hand, when 
\bea\label{condi1}
\ep_{X}\simeq \mathcal{O}(\ep^{2})
\eea
together with the condition in Eq.~(\ref{incomune}) we manage to include all $P(X,\phi)=f(X)+g(\phi)$ class of models (notice that Eq.~(\ref{condi1}) implies 
$\dot{\ep}_{X}\simeq \mathcal{O}(\ep^{3}),\quad\ddot{\ep}_{X}\simeq \mathcal{O}(\ep^{4})$). We report the results for all these cases in the second column of Table~\ref{ordini}. \\
Among the $P(X,\phi)=f(X)\times g(\phi)$ models, we list the orders of magnitude for the case described by Eq.~(\ref{ex}) in the third column of the table. Notice that, as shown in this section, the coefficients in Table 1 are not all independent from one another and can be combined in a smaller number of independent parameters, such as in Eqs.~(\ref{m}) through (\ref{o}) (see e.g. \cite{Chen:2006nt}). Another way of defining these parameters is presented in \cite{Cheung,SSZ} whithin an effective field theory framework, where the ones that are linearly independent are reported to leading order in slow-roll. \\
In our loop calculations, we will consider the leading order terms in the Lagrangian, restricting to the cases included in the first two columns of Table~\ref{ordini}. 
Notice that, taking the leading order terms only, i.e. neglecting the coefficients with derivatives w.r.t. $\phi$, allows to consider at the same time both class I and class II models on the same level.
\\

\section{$\delta$N formalism applied to $P(X,\phi)$ models}\label{delta n}

\noindent 
The $\delta$N formalism~\cite{deltaN1,deltaN2,deltaN3,deltaN4} relates the comoving curvature fluctuations defined on a uniform density temporal slice at time $t$ to the initial 
perturbations of all the fields defined on a flat slice at time $t^{*}<t$. In standard slow-roll inflation, we have   
\be\label{deltaN}
\ze(\vec{x},t)=\sum_{n}{\frac{N^{(n)}(t,t_*)}{n!}\left(\df(\vec{x})_*\right)^{n}},
\ee
where, $N^{(n)}(t,t_*)=\partial^{n} N/\partial \varphi_*^n$ represents the n-th derivative of the unperturbed number of e-foldings $N(t,t_*)$ 
w.r.t. the unperturbed values of the scalar field at time $t_*$. 
The computation of the power spectrum of the curvature perturbation produced during inflation can thus be made in two steps. First one needs to compute the power spectrum of the initial inflaton field 
perturbations $\delta \phi_*$, and then the $\delta$N formalism~(\ref{deltaN}) can be employed to obtain the final value of the power spcetrum at a later time t. 
As discussed in Refs.~\cite{deltaN3} and~\cite{Dimastrogiovanni:2008af}, even in the case where the tensor fluctuations are taken into account, they will not (explicitly) appear  
in the previous expansion. 
In the case of general single-field models of inflation it is convenient to choose the initial time $t_*$ a few e-folds after the crossing of the ``acoustic horizon'' which goes as $c_s \eta$.
A priori not only the field $\delta \phi$ but also its first time derivatives would in principle appear in the $\delta$N expansion, the reason being that the field $\phi$ obeys a second order differential equation and therefore its background trajectory can be uniquely determined once the field and its first derivatives are specified on the initial slice. 
In the case of standard single-field models, assuming the usual slow-roll approximation, 
such as $3 H\dot{\phi}\simeq - V^{'}(\phi)$, implies that $\phi$ and $\dot{\phi}$ are no more independent from each other, therefore the initial state for the system is completely determined by $\phi$ 
and it is not necessary to also include $\delta \dot{\phi}$ in (\ref{deltaN}). 

Since for the more general class of models with non-canonical kinetic terms in the 
Lagrangian $P(X,\phi)$ inflation can be attained even when the usual slow-roll parameters are not small, one could wonder whether such a 
simplification can be still employed, so as to still make use of the simple expression~(\ref{deltaN}). 
For $P(X,\phi)$ models, the equations of motion are in fact more complex. Nevertheless, the smallness of the slow-variation parameters and of $c_{s}$ allow to get rid of 
$\ddot{\phi}$ in (\ref{c}). Both for $P=f(X)+g(\phi)$ and for DBI models, we have indeed verified that (\ref{c}) reduces to
\bea\label{NB}
6HXP_{X}\simeq\sqrt{2X}P_{\phi},
\eea 
by taking into account the order of magnitude of the different terms appearing in Eq.~(\ref{c}) and the results of Sec.~\ref{beom} (see also \cite{Franche:2009gk} for another derivation of Eq.~(\ref{NB})). Like in the standard case, this is an equation relating $\phi$ and 
its first time derivative, which are, again, no more independent (it is easy to check that a similar result can be obtained for ``power-law'' K-inflation as in (\ref{ex})). In Ref.~\cite{Arroja:2008ga} there is a hint about the use of the $\delta N$ formula for general single-field models of inflation 
and the reader can deduce an alternative way which validates the use of Eq.~(\ref{deltaN}).\\

\noindent We will be using Eq.~(\ref{deltaN}) in the computation of the power spectrum $P_{\zeta}$ of the curvature fluctuations
\begin{equation}
\langle \zeta_{\vec{k}_1}(t) \zeta_{\vec{k}_2}(t) \rangle=(2\pi)^3 \delta^{(3)}(\vec{k}_1+\vec {k}_2) P_\zeta(k)\, .  
\end{equation}
Up to one-loop it can be expressed as \cite{Byrnes:2007tm,seery2}
\bea\label{QQQ}\fl
\langle \zeta_{\vec{k_{1}}}(t)\zeta_{\vec{k_{2}}}(t) \rangle&=&(2 \pi)^{3}\delta^{(3)}(\vec{k_{1}}+\vec{k_{2}})\Big[\left( N^{(1)} \right)^{2}\Big(P_{{\rm tree}}(k_1)
+P_{{\rm one-loop}}(k_1)\Big) \nonumber\\\fl
&+&N^{(1)}N^{(2)}\int \frac{d^{3}q}{(2 \pi)^{3}} B_{\phi}(k_1,q,|\vec{k}_1-\vec{q}|)\nonumber\\
&+&\frac{1}{2}\left( N^{(2)} \right)^{2}\int \frac{d^{3}q}{(2 \pi)^{3}} P_{{\rm tree}}(q)P_{{\rm tree}}(|\vec{k}_1-\vec{q}|)
\nonumber\\\fl
&+&N^{(1)}N^{(3)}P_{{\rm tree}}(k)\int \frac{d^{3}q}{(2 \pi)^{3}} P_{{\rm tree}}(q)\Big]\, .
\eea
In Eq.~(\ref{QQQ}) $P_{{\rm tree}}(k)$ is the tree level power spectrum of the inflaton field~\cite{Garriga:1999vw}
\bea\label{tree}
\langle\df_{\vec{k_{1}}}\df_{\vec{k_{2}}}\rangle_*&=&  (2 \pi)^{3}\delta^{(3)}(\vec{k_{1}}+\vec{k_{2}})
\frac{H_*^{2}}{2c_{s} P_{X}k^{3}} \, ,
\eea
with the Hubble parameter, and also the other quantities, evaluated at horizon exit (when $kc_s=a H$). $P_{{\rm one-loop}}(k)$ is the one-loop contribution to the powers spectrum of the initial 
inflaton fluctuations  
\begin{equation} 
P_{{\rm one-loop}}(k)=P_{{\rm scalar}}(k)+P_{{\rm tensor}}(k)\, ,
\end{equation}
where $P_{{\rm scalar}}$, accounts for the contributions coming from scalar loops, whereas $P_{{\rm tensor}}(k)$ is due to loop diagrams where tensor modes are involved. 
Both of these corrections will be computed in this paper.\\
\noindent Finally, the second line of (\ref{QQQ}) includes the integral of $B_{\phi}(k_{1},k_{2},k_{3})$, the bispectrum of the scalar field (see \cite{Chen:2006nt} for the evaluation of $B_{\zeta}$ for general $P(X,\phi)$ models).  
Since $\left(N^{(1)}\right)^2= (M_{P}^{-2})(P_{X}/2\ep)$ the tree-level power spectrum of the curvature perturbation reads 
\begin{equation}
P_\zeta(k)=\frac{H^2}{4 M^2_{p} k^3 \ep c_s}\, ,
\end{equation}
where all the quantities on the r.h.s. are meant to be evaluated at horizon-crossing. 
\\

\section{Perturbative expansion of the action}\label{perturbation-L}

The power spectrum of $\df$ can be computed using the Schwinger-Keldysh (also dubbed as ``in-in'') formula
\bea \label{eq3}\fl
\langle\Omega|\df_{\vec{k_{1}}}(\eta)\df_{\vec{k_{2}}}(\eta)|\Omega\rangle_{1L}&=&i\Big\langle 0\Big|T\left[\df_{\vec{k_{1}}}(\eta)\df_{\vec{k_{2}}}(\eta) \int^{\eta}_{- \infty}d \epr \left(H_{int}^{+}(\epr)-H_{int}^{-}(\epr)\right)\right]0\rangle\\
&+&\frac{(-i)^{2}}{2}\langle 0|T\Big[\df_{\vec{k_{1}}}(\eta)\df_{\vec{k_{2}}}(\eta) \int^{\eta}_{- \infty}d \epr \left(H_{int}^{+}(\epr)-H_{int}^{-}(\epr)\right)\nonumber\\\fl&\times&\int^{\eta}_{- \infty}d \eps \left(H_{int}^{+}(\eps)-H_{int}^{-}(\eps)\right)\Big]\Big|0\Big\rangle,\nonumber
\eea 
where $H_{int}$ is the interaction Hamiltonian and the plus and minus signs indicate the propagators in the in-in formalism \cite{in-in1,in-in2,in-in3} (see also \cite{weinberg1} for a complete review). \\
The field operators appearing on the right-hand side of Eq.(\ref{eq3}) can be expanded in terms of their Fourier modes
\bean
\df(\vec{x},\eta)=\int d^{3}k e^{i\vec{k}\vec{x}}\left[a_{\vec{k}} \df_{k}(\eta)+a^{+}_{-\vec{k}} \df_{k}^{*}(\eta)\right],\\
\ga_{ij}(\vec{x},t)=\int d^{3}k e^{i\vec{k}\vec{x}}\sum_{\lambda}{ \left[\ep_{ij}(\hat{k},\lambda)b_{\vec{k},\lambda}\ga_{k}(\eta)
+\ep^{*}_{ij}(-\hat{k},\lambda)b^{+}_{-\vec{k},\lambda}\ga^*_{k}(\eta)\right]},
\eean
with creation and annihilation operators given by
\bean
\left[a_{\vec{k}},a^{+}_{\vec{k'}}\right]=(2 \pi)^{3}\de^{(3)}(\vec{k}-\vec{k'}),\\
\left[b_{\vec{k},\lambda},b^{+}_{\vec{k'},\lambda^{'}}\right]=(2 \pi)^{3}\de^{(3)}(\vec{k}-\vec{k'})\de_{\lambda,\lambda^{'}}.
\eean
The eigenfunctions for scalar modes are \footnote{Notice that, at least for the cases we are taking into account, the prefactor $1/\sqrt{c_{s}P_{X}}$ in Eq.~(\ref{uk}) has a time derivative 
that is much smaller compared to the quantity itself (and even exactly equal to zero for the DBI models discussed in Sec.~\ref{examples}); this is important in order to be allowed to treat this prefactor as a constant when integrating 
over time. The same applies to the $(\dot{\phi})^{-n}$ ($n>0$) factors that appear in the interaction terms of the action (see for instance Eqs.~(\ref{s3})-(\ref{s4})).}   
\be\label{uk}
\delta\phi_{k}(\eta)=\frac{H}{\sqrt{2c_{s}P_{X} k^{3}}}\left(1+i k c_{s}\eta\right)e^{-ik c_{s}\eta},
\ee
whereas for the tensor modes
\be\label{gk}
\gamma_{k}(\eta)=\frac{2H}{\sqrt{2 k^{3}}}\left(1+i k \eta\right)e^{-ik\eta}.
\ee
One loop corrections to the power spectrum are sourced both by three and four-leg interactions, which means that the perturbative expansion of $H_{int}$ should be carried out up to fourth order in the field fluctuations. \\

\noindent It is convenient to adopt the $3+1$ Arnowitt-Deser-Misner (ADM) splitting for the metric tensor. 
In the spatially flat gauge the metric is
\bea\label{eq2}
ds^2=-N^2dt^2+h_{ij}(dx^{i}+N^{i}dt)(dx^{j}+N^{j}dt),\\
h_{ij}=a^{2}(t)(e^{\ga})_{ij},
\eea
where $a(t)$ is the unpertrubed scale factor, $\ga_{ij}$ is a tensor perturbation with $\p_{i}\ga_{ij}=\ga_{ii}=0$ (traceless and divergenceless) and det$(e^{\ga})_{ij}=1$. $N$ and $N_{i}$ are called ``lapse'' and ``shift'' functions and can be seen as Lagrange multipliers: they obey first order differential equations (respectively called the ``Hamiltonian'' and the ``momentum'' constraints) that can be solved to derive the expression of $N$ and $N_{i}$ as functions of the field perturbations $\delta\phi$ and $\gamma_{ij}$. The action (\ref{beginning}) can be written as follows \cite{Maldacena:2002vr}
\be\label{action-general}
S=\frac{1}{2}\int{dtd^{3}x \sqrt{h}\left[NR^{3}+2NP+N^{-1}\left(E_{ij}E^{ij}-E^{2}\right)\right]}
\ee
where
\bea
E_{ij}=\frac{1}{2}\left(\dot{h_{ij}}-\bt_{i}N_{j}-\bt_{j}N_{i}\right),\\
E=h^{ij}E_{ij},\\
R^{(3)}=-\frac{1}{4a^{2}}\p_{i}\gamma_{ab}\p_{i}\gamma_{ab}.
\eea
Following the general procedure of Refs.~\cite{Maldacena:2002vr,Chen:2006nt}, the next step is to write and solve the constraint equations, replace them in the action and expand the Lagrangian up to fourth order in the perturbations $\delta \phi$ and $\gamma_{ij}$. We solved the constraints in \cite{Dimastrogiovanni:2008af} (see also Appendix A of this paper for a review of the computation). The complete expressions for the third and fourth order action instead represent a new result and are provided in Appendix B. \footnote{A perturbative expansion of the Lagrangians for $P(X,\phi)$ models was done in \cite{Arroja:2008ga}, but with a different approach as far as the tensor modes are concerned; here we expand the tensor mode fluctuations in a similar way as we do for the scalar ones.}\\

\subsection{Leading terms in the third and fourth order actions}

Before we proceed with selecting the leading order terms according to the assumptions we made in Sec.~\ref{beom} (i.e. according to columns I and II of Table~\ref{ordini} therein), it is important to make a preliminary remark about the diagrams with tensor loops to fourth order. As shown in \cite{Dimastrogiovanni:2008af} and also pointed out in \cite{Adshead:2008gk}, the momentum integrand functions for these diagrams are expected to be independent of the external momentum $k$, therefore they can only exhibit renormalizable power-law ultraviolet divergences. Their final contribution should then be included in a leftover renormalization constant. For this reason, the computation of these contributions will not be performed. We have followed the same procedure to account for 
the one-vertex scalar loop diagrams (see below and Appendix D for a 
more detailed discussion of this point).\\
\noindent With this remark in mind, we will report in this section the action to third and fourth order. For the third order action, we report the leading order contributions from the scalar interactions and all of the contributions from the terms involving tensors (determining the order of magnitude of the tensor contributions is not as trivial as for the scalar ones, so we decide to report all of them at this point)
\bea\label{s3}\fl
S^{(3)}=\int d^{3}x d\eta \frac{a^{4}}{2}&\Bigg[&H\ga_{ij}\ga_{ik}\dot{\ga_{jk}}+\frac{1}{3}P_{XXX}\fd^3\ddf^3-\frac{P_{,XX}}{a^2}{\Big(\p_{i}\df\Big)}^{2}\fd\ddf-\frac{\dot{\phi}P_{X}}{8H}\dot{\gamma}_{ij}\dot{\gamma}_{ij}\df\nonumber\\ \fl
&+&\frac{\Sigma}{2H\fd}\dot{\ga_{ij}}\p_{k}\ga_{ij}\p_{k}\p^{-2}\delta\dot{\phi}
+\frac{P_{,X}}{8Ha^2}\Big(-\fd\df\p_{i}\ga_{jk}\p_{i}\ga_{jk}+8H\ga_{ij}\p_{i}\df\p_{j}\df\Big)\nonumber\\ \fl&-&\frac{P_{,X}\Sigma}{2H^2}\dot{\ga_{ij}}\df\p_{i}\p_{j}\p^{-2}\ddf-\frac{\Sigma^2 }{\fd^2 H^2}\p_{i}\ga_{jk}\p_{k}\p^{-2}\ddf\p_{i}\p_{j}\p^{-2}\delta\dot{\phi}
\Bigg].\nonumber\\
\eea
\noindent For the fourth order action, we report the leading order scalar terms only 
\bea\label{s4}\fl
S^{(4)}=\int d^{3}x d\eta \frac{a^{4}}{2}&\Bigg[&\frac{1}{12}P_{,XXXX}\fd^4\ddf^4+\frac{12\lambda\Sigma}{\fd^4 H}\tilde{\alt}\p^{2}\tilde{\tht}-\frac{24\lambda\Sigma}{\fd^4 H}\left(\p_{i}\p_{j}\tilde{\tht}\right)\left(\p_{i}\tilde{\b_{j}}\right)\nonumber\\
\fl
&+&P_{,X}\Bigg(-\frac{P_{,XXXX}\fd^6}{6H}\df\ddf^3+\frac{P_{,XXX}\fd^4}{2Ha^2}\df\ddf\left(\p_{i}\df\right)^2\Bigg)\nonumber\\
\fl&-&\frac{P_{XXX}\Sigma\dot{\phi}^2}{H}\tilde{\alt}\ddf^2+\frac{P_{,XX}}{H}\Bigg(\frac{\Sigma}{a^2}\left(\p_{i}\df\right)^2\tilde{\alt}+6
\lambda\ddf\p_{i}\df\p_{i}\tilde{\tht}\nonumber\\
\fl&+&\frac{H}{4a^4}{\Big(\p_{i}\df\Big)}^{4}+\frac{\Sigma}{a^2}\left(\p_{i}\p^{-2}\ddf\right)\left(\p_{i}\df\right)\left(\p_{j}\df\right)^2\Bigg)
-\frac{9\lambda^2}{\dot{\phi}^4 H^2}\delta\dot{\phi}^4\nonumber\\
\fl&-&\frac{\dot{\phi}^2P_{XXX}}{2a^2}\delta\dot{\phi}^2\left(\p_{i}\delta\phi\right)^2+\frac{9\lambda^2}{\dot{\phi}^4 H^2}\left(\p_{i}\p_{j}\tilde{\theta}_{2}\right)^2   \Bigg].
\eea
We indicate $\left(\p_{i}\df\right)^2\equiv \delta_{ij}\p_{i}\df\p_{j}\df$ and sums are intended over repeated spatial indices. Also, we define
\bea
\tilde{\tht}\equiv\p^{-2}\left(\ddf\right)^2 \\
\tilde{\alt}\equiv \p^{-2}\left[A+B\right]\\
\tilde{\beta_{j}}\equiv \p^{-2}\left[\p^{-2}\p_{j}\left(A+B\right)-C\right]
\eea
where $A \equiv\left(\p^{2}\df\right)\ddf$, $B \equiv\delta_{ij}\left(\p_{i}\df\right)\left(\p_{j}\ddf\right)$ and $C\equiv \left(\p_{j}\df\right)\ddf$.\\

\noindent The complete (i.e. without any slow-variation parameter approximation and including all tensor mode interactions also at fourth order) expression of the action up to fourth order were computed and can be found in Appendix B. It is important to point out that the selection criteria we adopted in order to identify the leading scalar contributions, although they apply to a large number of $P(X,\phi)$ models (as described in Sec.~\ref{beom}), are not valid for all of them. However, our complete expression of the interaction Lagrangian offers the possibility of computing loop corrections (as well as higher order correlators) in $P(X,\phi)$ models other than the ones we considered.  \\

\subsection{Calculation of the interaction hamiltonian}\label{abbreviations}

In a theory, like the one we are dealing with, where time derivatives of the fields appear in the non-quadratic terms of the Lagrangian, the relation $H_{int}=-L_{I}$ between the interaction Hamiltonian and the Lagrangian does not generally hold \cite{peskin,seery1,Huang:2006eha} (also see \cite{Dimastrogiovanni:2008af} for a quick discussion on this). \\
We follow the prescription reviewed in \cite{Adshead:2008gk} for computing $H_{int}$. It is convenient to schematically write the Lagrangian up to fourth order as follows 
\bea\label{ih}\fl
\textit{L}&=&\Big[f_{0}^{(\df)}\ddf^2+j_{2}^{(\df)}+f_{0}^{(\ga)}\dot{\ga}^2+j_{2}^{(\ga)}+g_{0}^{(\df)}\ddf^3+g_{2}^{(\df)}\ddf+g_{2}^{(\ga)}\dot{\ga}+g_{1}^{(\ga)}\dot{\ga}\ddf+g_{1}^{(\df)}\dot{\gamma}^2\nonumber\\ \fl&+&G_{1}^{(\ga)}\ddf^2+j_{3}^{(\ga,\ga,\df)}+j_{3}^{(\ga\df\df)}+h_{0}^{(\df)}\ddf^4+h^{(\df)}_{1}\ddf^3+h^{(\df)}_{2}\ddf^2+h^{(\df)}_{3}\ddf+j_{4}^{(\df)}\Big]\, ,
\eea
where $f$, $g$, $G$, $j$ and $h$ are functions of the arguments (and their derivatives) that are indicated in their upper indices, 
whereas the lower index refers to their pertubation order. They can be easily read from Eqs.~(\ref{s3}) and~(\ref{s4}). Notice that we have for simplicity ignored the tensor indices.\\ 
The momentum densities for the scalar and tensor perturbations are
\bea
\pi^{(\df)}&\equiv& \frac{\p \textit{L}}{\p \ddf}\, ,\\
\pi^{(\ga)} &\equiv& \frac{\p \textit{L}}{\p \dot{\ga}}\, ,
\eea
from which we have
\bea\label{1}
\ddf=\ddf^{(1)}+\ddf^{(2)}+\ddf^{(3)}\, ,
\eea
where
\bea\fl
\ddf^{(1)}&=&\frac{\pi^{(\df)}}{2f_{0}^{(\df)}}\\, ,\
\fl
\ddf^{(2)}&=&-\frac{g_{2}^{(\df)}}{2f_{0}^{(\df)}}-\frac{3g_{0}^{(\df)}{\pi^{(\df)}}^{2}}{{\Big(2f_{0}^{(\df)}\Big)}^{3}}-\frac{G_{1}^{(\ga)}\pi^{(\df)}}{{2\Big (f_{0}^{(\df)}\Big)}^{2}}-\frac{g_{1}^{(\ga)}\pi^{(\ga)}}{4f_{0}^{(\df)}f_{0}^{(\ga)}}\, , \\
\fl
\ddf^{(3)}&=&\frac{9 \left(g_{0}^{(\df)}\right)^2\pi^{(\df)}}{16 \left(f_{0}^{(\df)}\right)^5}+\frac{3g_{0}^{(\df)}g_{2}^{(\df)}\pi^{(\df)}}{4\left(f_{0}^{(\df)}\right)^3}-\frac{h_{0}^{(\df)}\left(\pi^{(\df)}\right)^3}{4\left(f_{0}^{(\df)}\right)^4}-\frac{h_{2}^{(\df)}\pi^{(\df)}}{2\left(f_{0}^{(\df)}\right)^2}-\frac{h_{3}^{(\df)}}{2 f_{0}^{(\df)}}\nonumber\\\fl&-&\frac{3 h_{1}^{(\df)}\left(\pi^{(\df)}\right)^2}{8\left(f_{0}^{(\df)}\right)^3}-\frac{g_{1}^{(\ga)}\dot{\ga}^{(2)}}{2f_{0}^{(\df)}}-\frac{G_{1}^{(\ga)}\delta\dot{\phi}^{(2)}}{f_{0}^{(\df)}}\, ,
\eea
and
\bea\label{2}
\dot{\ga}=\dot{\ga}^{(1)}+\dot{\ga}^{(2)}+\dot{\ga}^{(3)}\, ,
\eea
where
\bea
\dot{\ga}^{(1)}=\frac{\pi^{(\ga)}}{2f_{0}^{(\ga)}}\, ,\\
\dot{\ga}^{(2)}=-\frac{g_{2}^{(\ga)}}{2f_{0}^{(\ga)}}-\frac{g_{1}^{(\ga)}\pi^{(\df)}}{4f^{(\df)}_{0}f^{(\ga)}_{0}}-\frac{g_{1}^{(\df)}\pi^{(\ga)}}{2\left(f^{(\ga)}_{0}\right)^2} \, ,\\
\dot{\ga}^{(3)}=-\frac{g_{1}^{(\ga)}\delta\dot{\phi}^{(2)}}{2f_{0}^{(\ga)}}-\frac{g_{1}^{(\df)}\dot{\ga}^{(2)}}{f_{0}^{(\ga)}}\, .
\eea
The Hamiltonian density 
\be
\textit{H}=\pi^{(\df)}\ddf+\pi^{(\ga)}\dot{\ga}-\textit{L}
\ee
is derived plugging in Eqs.\,(\ref{1}) and (\ref{2}) into its definition. We finally replace the conjugate momenta with the time derivatives of the field perturbations as derived from the free Hamiltonian density $\textit{H}_{0}$
\bea
\delta\dot\phi &\equiv& \frac{\p \textit{H}_{0}}{\p \pi^{(\df)}}\\
\dot\ga &\equiv& \frac{\p \textit{H}_{0}}{\p \pi^{(\ga)}}
\eea
and the result is
\bea\label{3}\fl
\textit{H}_{int}^{(3)}&=&-\textit{L}_{int}^{(3)}
\eea
for the third order part of the interaction Hamiltonian and
\bea\fl
\label{4}
\textit{H}_{int}^{(4)}&=&\frac{\delta\dot{\phi}^4}{4}\left(\frac{9g_{0}^{(\df)}}{f_{0}^{(\df)}}-4h_{0}^{(\df)}\right)-h_{1}^{(\df)}\delta\dot{\phi}^3+\frac{\delta\dot{\phi}^2}{2}\left(\frac{3g_{0}^{(\df)}g_{2}^{(\df)}}{f_{0}^{(\df)}}-2h_{2}^{(\df)}\right)-h_{3}^{(\df)}\delta\dot{\phi}\nonumber\\\fl&+&\frac{\left(g_{2}^{(\df)}\right)^2}{4f_{0}^{(\df)}}-j_{4}^{(\df)}+\frac{\dot{\ga}^2}{4}\left(\frac{4\left(g_{1}^{(\df)}\right)^2}{f_{0}^{(\ga)}}+\frac{\left(g_{1}^{(\ga)}\right)^2}{f_{0}^{(\df)}}\right)\nonumber\\\fl&+&\frac{\dot{\ga}}{2f_{0}^{(\df)}f_{0}^{(\ga)}}\left(f_{0}^{(\ga)}g_{1}^{(\ga)}\left(3\delta\dot{\phi}^2 g_{0}^{(\df)}+2\delta\dot{\phi}G_{1}^{(\ga)}+2g_{2}^{(\df)}\right)+2f_{0}^{(\df)}g_{1}^{(\df)}\left(\delta\dot{\phi}g_{1}^{(\ga)}+g_{2}^{(\ga)}\right)\right)\nonumber\\\fl&+&\frac{4f_{0}^{(\ga)}G_{1}^{(\ga)}\delta\dot{\phi}\left(3g_{0}^{(\df)}\delta\dot{\phi}^2+G_{1}^{(\ga)}\delta\dot{\phi}+g_{2}^{(\df)}\right)+f_{0}^{(\df)}\left(g_{1}^{(\ga)}\delta\dot{\phi}+g_{2}^{(\ga)}\right)^2}{4f_{0}^{(\df)}f_{0}^{(\ga)}}
\eea
for the fourth order part. In the case where the tensor perturbations are set to zero one can verify that Eqs.\,(\ref{3}) and (\ref{4}) reproduce Eq.(13) and (14) in \cite{Huang:2006eha}.\\

\section{Calculation of the loop corrections}\label{actualcomp}

In this section we will provide the order of magnitude of the one-loop corrections arising from the third and the fourth order interaction Hamiltonian. \\
\noindent The order of magnitude of the different diagrams can be quickly estimated for the diagrams with scalar loops. In fact, for these diagrams it is possible to factor out of the time and of the momentum integrals all of the $H$, $c_{s}$, $P_{X}$, $\ep$ and all other parameters of the theory. For the diagrams with tensor loops it is instead necessary to evaluate the time integrals before the overall coefficient can be determined. We first explain the prescription for deriving the order of magnitude of the final result for each type of diagram and, after that, we provide the actual orders of magnitude for all the leading order diagrams. We 
report their precise expression in Sec.~\ref{finalresults}.\\

\begin{itemize}

\item{Scalar loop diagrams}\\

\begin{itemize}
\item \textit{Two-vertex diagrams}

Let us consider two three-scalar interaction verteces, which we will indicate as $V_{1}$ and $V_{2}$, from the interaction Hamiltonian (i.e. $\int d^3x d\eta a^{4}H_{int}^{(3)}\supset\int d^3x d\eta a^{4}(\eta)(V_{1}+V_{2})$). The verteces $V_{i}$ ($i=1,2$) will in general be expressed as $V_{i}\equiv v_{i}(H,\ep,P_{X},...)f_{i}(\df,\p_{\mu}\df)$, i.e. as the product of a function of some parameters of the theory, multiplied by the scalar field fluctuations and their derivatives. The prescription for calculating the overall (dimensionless) coefficient $\mathcal{C}$ of the diagram built from $V_{1}$ and $V_{2}$ (as in Fig.~\ref{2v-scal}) is the following
\bea\label{spiego} 
\mathcal{C}\simeq(v_{1})\times(v_{2})\times\left(\frac{H}{\sqrt{c_{s}P_{X}}}\right)^{8}\times\left(\frac{c_{s}^3}{H^4}\right)^2\times(H)^{\sharp time\,der},
\eea
the complete analytic expression of the diagram being given by 
\bea\label{spiego0}
\langle\df_{\vec{k}_{1}}(\eta^{*})\df_{\vec{k}_{2}}(\eta^{*})\rangle\supset\delta^{(3)}(\vec{k}_{1}+\vec{k}_{2})\mathcal{C}G(x^{*})F(k), 
\eea
where $F$ is a function of the external momentum $k\equiv |\vec{k}_{1}|=|\vec{k}_{2}|$ with dimensions (mass)$^{-3}$ and $G$ is a function of $x^{*}\equiv-kc_{s}\eta^{*}$. The coefficient $(H/(\sqrt{c_{s}P_{X}}))^8$ comes from having eight eigenfunctions like the one in Eq.~(\ref{uk}); $(c_{s}^3/H^4)^2$ comes from having two temporal integrations $d\eta a^{4}(\eta)$, which we rewrite in terms of the dimensionless variable $x$ as $(dx/x^4)(-k^3c_{s}^{3}/H^4)$, using $a\simeq(H\eta)^{-1}$, valid for a quasi de-Sitter spacetime; finally, extra factors of $H$ are needed for each temporal derivative (spatial derivatives do not contribute to determine $\mathcal{C}$) appearing in the vertices, as we can see from the following expression
\bea\label{qui}\fl
\delta\dot{\phi}_{k}(\eta)=-\frac{H^2}{\sqrt{2c_{s}P_{X} k^{3}}}x^2e^{ix}.
\eea
\\
\item{One-vertex diagrams}

If we consider diagrams with only one vertex $V\equiv v(H,\ep,P_{X},...)f(\df,\p_{\mu}\df)$ arising from scalar interactions (see Fig.~\ref{1v-scal}), the prescription will be slightly different
\bea\label{spiego1} 
\mathcal{C}\simeq(v)\times\left(\frac{H}{\sqrt{c_{s}P_{X}}}\right)^{6}\times\left(\frac{c_{s}^3}{H^4}\right)\times(H)^{\sharp time\,der},
\eea
given that the number of time integrals has now gone from two to one and the number of eigenfunctions from eight to six. \\

%The leading order vertices containing scalar field perturbations only are (from the previous sections) $V_{a}\equiv a^{4}g_{0}^{(\df)}\ddf^3$ and $V_{b} \equiv a^{4}g_{2}^{(\df)}\ddf$.\\
%They produce the following diagrams (we will indicate the order of magnitude of the diagram on the right hand side whereas $D_{(ij)}$ stands for the diagrams generated by the vertices $i$ and $j$):
%\be\label{partial1}\fl
%D_{\left(V_{a}V_{a}\right)}\sim D_{\left(V_{a}V_{b}\right)} \sim D_{\left(V_{b}V_{b}\right)}\sim \frac{\ep^2}{c_{s}^{6}\fd^6}=\frac{P_{,X}^{3}}{\ep c_{s}^{6}}
%\ee

\end{itemize}

\begin{figure}
\begin{center}
\scalebox{0.5}{\includegraphics{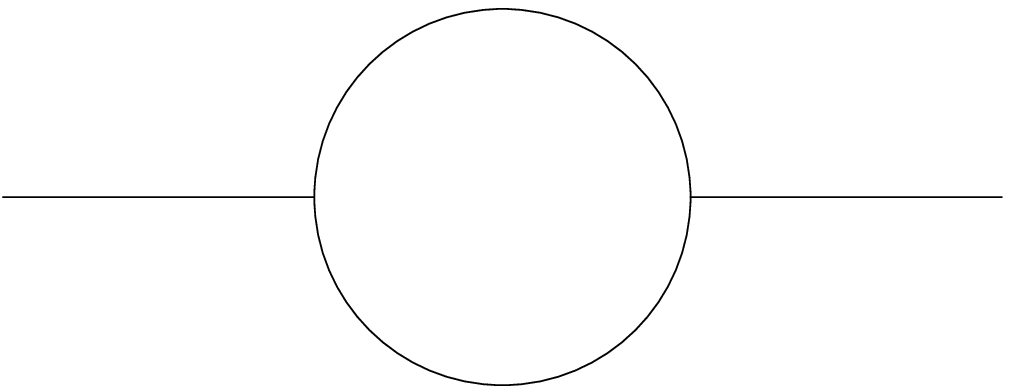}} 
\caption{One-loop two-vertex corrections to $P_{\delta\phi}$ from scalar modes.}
\label{2v-scal}
\end{center}
\end{figure}

\begin{figure}
\begin{center}
\scalebox{0.5}{\includegraphics{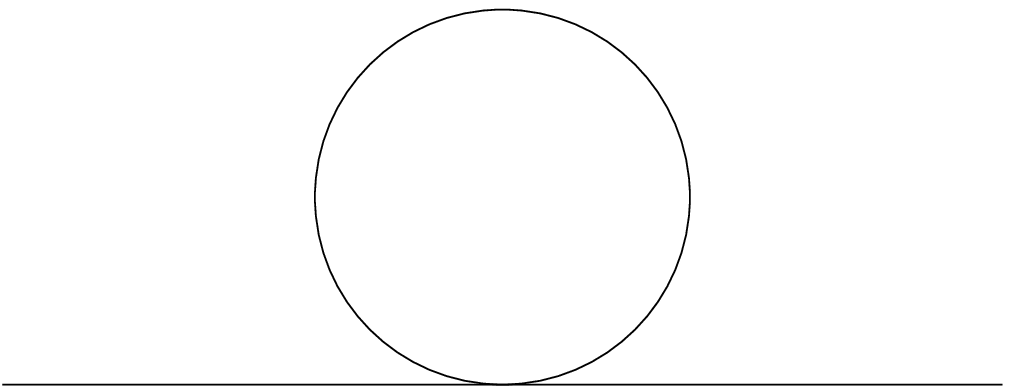}} 
\caption{One-loop one-vertex corrections to $P_{\delta\phi}$ from scalar modes.}
\label{1v-scal}
\end{center}
\end{figure}

\begin{figure}
\begin{center}
\scalebox{0.5}{\includegraphics{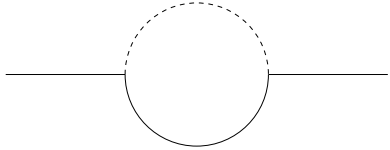}}
\caption{One-loop tensor corrections to $P_{\delta\phi}$ from two-scalar one-graviton ($\sim\gamma\left(\delta\phi\right)^2$) interactions. The continuous line represent a scalar propagator, the dotted line a tensor propagator.}
\label{2v-tens}
\end{center}
\end{figure}

\begin{figure}
\begin{center}
\scalebox{0.5}{\includegraphics{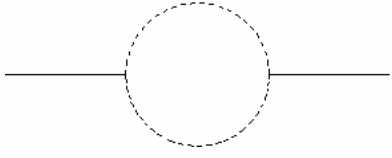}}
\caption{One-loop tensor corrections to $P_{\delta\phi}$ from two-graviton one-scalar ($\sim\delta\phi\left(\gamma\right)^2$) interactions.}
\label{2v-tens1}
\end{center}
\end{figure}

\item{Tensor loop diagrams}\\

The evaluation of $\mathcal{C}$ for diagrams with tensor modes running in the loop is not as straightforward as for the scalar diagrams because the sound speed cannot be so easily factored out before actually solving the time integrals. In fact, the time integrals will be over a combination of derivatives of the eigenfunctions of the scalar modes from Eq.~(\ref{qui}) and of the eigenfunctions of the tensor modes
\bea\label{qua}
\dot{\gamma}_{k}(\eta)=-\frac{2H}{\sqrt{2 k^{3}}}\frac{Hx^2}{c_{s}^2}e^{i\frac{x}{c_{s}}}.
\eea
However we expect that the one-loop corrections involving the tensor perturbations will be subdominat with respect to pure scalar contributions, the reason being 
that the scalar field fluctuations depend on the sound speed, while the tensor ones do not, and the order of the loop corrections is dominated by inverse powers of 
the sound speed. Such an expectation is indeed confirmed by the estimates below.

\begin{itemize}

\item \textit{Two-scalar one-graviton interactions}

If the vertices $V_{1}$ and $V_{2}$ indicate an interaction between two scalar and one tensor degrees of freedom (see Fig.~\ref{2v-tens}), the prescription for the $\mathcal{C}$ coefficient becomes
\bea\label{spiego2}\fl 
\mathcal{C}\simeq(v_{1})\times(v_{2})\times\left(\frac{H}{\sqrt{c_{s}P_{X}}}\right)^{6}\times(H)^{2}\times\left(\frac{c_{s}^3}{H^4}\right)^2\times(H)^{\sharp time\,der}\times\left(\int dxdx\right)\nonumber\\
\eea
where the $H^2$ factor comes from having two tensor eigenfunctions in the diagram and $\left(\int dxdx\right)$ reminds us that extra $c_{s}$ factors are 
expected from the time integrals.

\item \textit{Two-graviton one-scalar interactions}

Finally, for diagrams as in Fig.~\ref{2v-tens1}, we have
\bea\label{spiego3}\fl
\mathcal{C}\simeq(v_{1})\times(v_{2})\times\left(\frac{H}{\sqrt{c_{s}P_{X}}}\right)^{4}\times(H)^{4}\times\left(\frac{c_{s}^3}{H^4}\right)^2\times(H)^{\sharp time\,der}\times\left(\int dxdx\right).\nonumber\\
\eea

\end{itemize}

\end{itemize}

\subsection{Order of magnitude of all leading order diagrams}

We are now ready to list the orders of magnitude of our diagrams. \\
Let us begin with the scalar loop diagrams. The leading order three-scalar interaction Hamiltonian terms are $\int d^3x d\eta a^4 H_{int}^{(3)}\supset\int d^3x d\eta a^4 \left[V_{a}+V_{b}\right]$, where 
\bea
V_{a}=- g_{0}^{(\df)}\ddf^3,\\
V_{b}= -g_{2}^{(\df)}\ddf, 
\eea
i.e. (from Eq.~(\ref{s3}))
\bea
V_{a}=-\frac{1}{3}P_{XXX}\dot{\phi}^3\ddf^3,\\
V_{b}=\frac{P_{XX}\dot{\phi}}{a^2}\ddf\left(\p_{i}\df\right)^2. 
\eea
From the combination of $V_{a}$ and $V_{b}$ we get three one-loop diagrams with similar orders of magnitude in terms of the parameters of the theory
\bea\label{leading}
\langle\df_{\vec{k}_{1}}(\eta^{*})\df_{\vec{k}_{2}}(\eta^{*})\rangle_{ij}=\delta^{(3)}(\vec{k}_{1}+\vec{k}_{2})\frac{H^{4}_{*}}{M_{P}^2}\frac{1}{\ep P_{X}c_{s}^{6}}G_{ij}(x^{*})F_{ij}(k),
\eea
where $i,j=a,b$ and, as usual, the $F_{ij}$ are functions of the external momentum with dimensions (mass)$^{-3}$ (here we reintroduce the Planck mass), while $G_{ij}$ are functions 
of $x_*=-k c_s \eta_*$. 
The explicit computation of the corresponding one-loop corrections to the power spectrum of $\delta \phi$ reads 
\bea\label{risf}\fl
\langle\df_{\vec{k}_{1}}(\eta^{*})\df_{\vec{k}_{2}}(\eta^{*})\rangle &\supset& \delta^{(3)}(\vec{k}_{1}+\vec{k}_{2})\pi\Bigg[\frac{8}{9}\left(X^3P_{XXX}\right)^2\frac{c_{s}^{2}}{\ep^3 P_{X}M_{P}^{6}}\left(\frac{G_{aa}(x^{*})\ln\left(\Lambda/H_{*}\right)}{k^3}\right)\nonumber\\\fl&-&\left(\frac{8}{3}\right)\left(X^{2}P_{XX}\right)\left(X^{3}P_{XXX}\right)\frac{1}{\ep^3 P_{X}M_{P}^{6}}\left(\frac{G_{ab}(x^{*})\ln\left(\Lambda/H_{*}\right)}{k^3}\right)\nonumber \\\fl&+&2\left(X^{2}P_{XX}\right)^2\frac{1}{\ep^3 c_{s}^{2} P_{X}M_{P}^{6}}\left(\frac{G_{bb}(x^{*})\ln\left(\Lambda/H_{*}\right)}{k^3}\right)\Bigg]
\eea
where $G_{aa}$, $G_{ab}$ and $G_{bb}$ are functions of $x^{*}$ as provided in Eqs.~(\ref{fr1}) through (\ref{fr3}) and $\Lambda$ 
is a fixed physical cutoff that was introduced while integrating over momentum \cite{SZnew}.\\

The result in Eq.~(\ref{risf}) can be rewritten after expressing $P_{XX}$ and $P_{XXX}$ in terms of the slow-variation parameters 
and in terms of the sound speed, according to Table~1. From Eq.~(\ref{qui1}) we have
\bea\label{repl1}
X^2 P_{XX}=\frac{\ep H^2 M_{P}^{2}}{2}\left(\frac{1}{c_{s}^{2}}-1\right)\simeq\frac{\ep H^2M_{P}^{2}}{2c_{s}^{2}}
\eea
where, in the last step, we have as usual assumed that we are in the regime of small sound speed. From Eq.~(\ref{qui2}) and for models 
which satisfy the constraint $Xc_{s}c_{sX}\sim\mathcal{O}(1)$, we have
\bea\label{repl2}\fl
X^3 P_{XXX}\simeq \frac{\ep H^2M_{P}^{2}}{c_{s}^{4}}.
\eea
Using (\ref{repl1}) and (\ref{repl2}), Eq.~(\ref{risf}) becomes
\bea\label{risf1}\fl
\langle\df_{\vec{k}_{1}}(\eta^{*})\df_{\vec{k}_{2}}(\eta^{*})\rangle \supset (2\pi)^3\delta^{(3)}(\vec{k}_{1}+\vec{k}_{2})\frac{H^4}{8\pi^2\ep c_{s}^{6} P_{X}M_{P}^{2}}\Bigg[\left(\frac{8}{9}G_{aa}-\frac{8}{6}G_{ab}+\frac{1}{2}G_{bb}\right)\frac{\ln\left(\Lambda/H_{*}\right)}{k^3}\Bigg].\nonumber\\
\eea

From the fourth order action, the leading order vertices are 
\bea
H_{int}^{(4)}\supset\int d  \eta d^{3}x a^{4}(t)\left(V_{(0)}+V_{(1)}+V_{(2)}+V_{(3)}+V_{(4)}\right), 
\eea
where
\bea\label{vertice01}
V_{(0)}\equiv \frac{P_{XXX}\Sigma\fd^2}{H}\tilde{\alt}\ddf^2,\\\label{vertice11}
V_{(1)}\equiv 24\frac{\lambda \Sigma}{H\fd^4}\left(\p_{i}\p_{j}\tilde{\tht}\right)\left(\p_{i}\tilde{\beta}_{j}\right),\\\label{vertice21}
V_{(2)}\equiv -12\frac{\lambda\Sigma}{H\fd^4}\tilde{\alt}\ddf^2,\\\label{vertice31}
V_{(3)}\equiv -\frac{P_{XX}\Sigma}{a^2 H}\tilde{\alt}\left(\delta_{ij}\p_{i}\df\p_{j}\df\right),\\\label{vertice41}
V_{(4)}\equiv -6\frac{P_{XX}\lambda}{H}\ddf\delta_{ij}\left(\p_{i}\df\right)\left(\p_{j}\tilde{\tht}\right).
\eea
Notice that the leading order vertices in $H^{(4)}_{int}$ from Eq.~(\ref{4}) (e.g. $\sim h_{0}^{(\df)}\ddf^4$) have not been taken into
account since they provide disconnected-like contributions, similarly to what
happens for the one-vertex diagrams with graviton loops (see also Appendix D
for an accurate discussion on one-vertex diagrams).\\
Using the abbreviations of Sec.~\ref{abbreviations}, these terms can be condensed as 
\bea
\textit{H}_{int}^{(4)}=-h^{(\df)}_{1}\ddf^3-h_{3}^{(\df)}\ddf.
\eea
\\
The order of magnitude of the one-loop diagrams built from the vertices (\ref{vertice01}) through (\ref{vertice41}) is given by
\bea
\langle\df_{\vec{k}_{1}}(\eta^{*})\df_{\vec{k}_{2}}(\eta^{*})\rangle_{i}=\delta^{(3)}(\vec{k}_{1}+\vec{k}_{2})\frac{H^{4}_{*}}{M_{P}^2}\frac{1}{P_{X}c_{s}^{6}}G_{i}(x^{*})F_{i}(k)
\eea
where now $i=0,1,...,4$. These are evidently subleading compared to (\ref{leading}).\\

\noindent Let us now move to the diagrams with tensor loops. The interaction terms with two gravitons and one scalar are
\bea
V_{g_{1}}=-\frac{\Sigma}{2H\fd}\dot{\ga_{ij}}\p_{k}\ga_{ij}\p_{k}\p^{-2}\ddf,\\
%V_{g_{2}}= -\frac{\Sigma}{\fd}\ga_{ij}\ga_{jk}\p_{i}\p_{k}\p^{-2}\ddf ,\\
V_{g_{2}}= \frac{\fd P_{,X}}{8Ha^2}\df\p_{i}\ga_{jk}\p_{i}\ga_{jk} ,\\
V_{g_{3}}= \frac{\dot{\phi}P_{X}}{8 H}\dot{\ga}_{ij}\dot{\ga}_{ij}\df .
\eea
The orders of magnitude of the one loop diagrams resulting from their combinations are as follows
\bea\fl
\langle\df_{\vec{k}_{1}}(\eta^{*})\df_{\vec{k}_{2}}(\eta^{*})\rangle_{g_{1}g_{1}}=\delta^{(3)}(\vec{k}_{1}+\vec{k}_{2})\frac{H^{4}_{*}}{M_{P}^2}\frac{\ep}{P_{X}c_{s}^{5}}G_{g_{1}g_{1}}(x^{*})F_{g_{1}g_{1}}(k),\\\fl
%\langle\df_{\vec{k}_{1}}(\eta^{*})\df_{\vec{k}_{2}}(\eta^{*})\rangle_{g_{1}g_{2}}=\delta^{(3)}(\vec{k}_{1}+\vec{k}_{2})\frac{H^{4}_{*}}{m_{P}^4}\frac{\ep}{P_{X}c_{s}^{4}}G_{g_{1}g_{2}}(x^{*})F_{g_{1}g_{2}}(k),\\\fl
\langle\df_{\vec{k}_{1}}(\eta^{*})\df_{\vec{k}_{2}}(\eta^{*})\rangle_{g_{1}g_{2}}=\delta^{(3)}(\vec{k}_{1}+\vec{k}_{2})\frac{H^{4}_{*}}{M_{P}^2}\frac{\ep}{P_{X}c_{s}^{4}}G_{g_{1}g_{2}}(x^{*})F_{g_{1}g_{2}}(k),\\\fl
\langle\df_{\vec{k}_{1}}(\eta^{*})\df_{\vec{k}_{2}}(\eta^{*})\rangle_{g_{1}g_{3}}=\delta^{(3)}(\vec{k}_{1}+\vec{k}_{2})\frac{H^{4}_{*}}{M_{P}^2}\frac{\ep}{P_{X}c_{s}^{4}}G_{g_{1}g_{3}}(x^{*})F_{g_{1}g_{3}}(k),\\\fl
%\langle\df_{\vec{k}_{1}}(\eta^{*})\df_{\vec{k}_{2}}(\eta^{*})\rangle_{g_{2}g_{2}}=\delta^{(3)}(\vec{k}_{1}+\vec{k}_{2})\frac{H^{4}_{*}}{m_{P}^4}\frac{\ep}{P_{X}c_{s}^{3}}G_{g_{2}g_{2}}(x^{*})F_{g_{2}g_{2}}(k),\\\fl
%\langle\df_{\vec{k}_{1}}(\eta^{*})\df_{\vec{k}_{2}}(\eta^{*})\rangle_{g_{2}g_{3}}=\delta^{(3)}(\vec{k}_{1}+\vec{k}_{2})\frac{H^{4}_{*}}{m_{P}^4}\frac{\ep}{P_{X}c_{s}^{3}}G_{g_{2}g_{3}}(x^{*})F_{g_{2}g_{3}}(k),\\\fl
%\langle\df_{\vec{k}_{1}}(\eta^{*})\df_{\vec{k}_{2}}(\eta^{*})\rangle_{g_{2}g_{4}}=\delta^{(3)}(\vec{k}_{1}+\vec{k}_{2})\frac{H^{4}_{*}}{m_{P}^4}\frac{\ep}{P_{X}c_{s}^{3}}G_{g_{2}g_{4}}(x^{*})F_{g_{2}g_{4}}(k),\\\fl
\langle\df_{\vec{k}_{1}}(\eta^{*})\df_{\vec{k}_{2}}(\eta^{*})\rangle_{g_{2}g_{2}}=\delta^{(3)}(\vec{k}_{1}+\vec{k}_{2})\frac{H^{4}_{*}}{M_{P}^2}\frac{\ep}{P_{X}c_{s}^{3}}G_{g_{2}g_{2}}(x^{*})F_{g_{2}g_{2}}(k),\\\fl
\langle\df_{\vec{k}_{1}}(\eta^{*})\df_{\vec{k}_{2}}(\eta^{*})\rangle_{g_{2}g_{3}}=\delta^{(3)}(\vec{k}_{1}+\vec{k}_{2})\frac{H^{4}_{*}}{M_{P}^2}\frac{\ep}{P_{X}c_{s}^{3}}G_{g_{2}g_{3}}(x^{*})F_{g_{2}g_{3}}(k)\\\fl
\langle\df_{\vec{k}_{1}}(\eta^{*})\df_{\vec{k}_{2}}(\eta^{*})\rangle_{g_{3}g_{3}}=\delta^{(3)}(\vec{k}_{1}+\vec{k}_{2})\frac{H^{4}_{*}}{M_{P}^2}\frac{\ep}{P_{X}c_{s}^{3}}G_{g_{3}g_{3}}(x^{*})F_{g_{3}g_{3}}(k),
\eea
which were computed after integrating in time.\\

\noindent Finally, the interaction terms with two scalars and one graviton are
\bea
V_{s_{1}}= -\frac{P_{,X}}{a^2} \left(\ga_{ij}\p_{i}\df\p_{j}\df\right),\\
V_{s_{2}}=\frac{P_{,X}\Sigma}{2H^2}\left(\dot{\ga_{ij}}\df\p_{i}\p_{j}\p^{-2}\ddf\right) ,\\
V_{s_{3}}=\frac{\Sigma^2 }{\fd^2 H^2}\left(\p_{i}\p_{j}\p^{-2}\ddf\right)\left(\p_{i}\ga_{jk}\right)\p_{k}\p^{-2}\ddf.
\eea
From the combination of these last vertices, we have the following one-loop corrections
\bea\fl
\langle\df_{\vec{k}_{1}}(\eta^{*})\df_{\vec{k}_{2}}(\eta^{*})\rangle_{s_{1}s_{1}}=\delta^{(3)}(\vec{k}_{1}+\vec{k}_{2})\frac{H^{4}_{*}}{M_{P}^2}\frac{1}{P_{X}c_{s}^{2}}G_{s_{1}s_{1}}(x^{*})F_{s_{1}s_{1}}(k),\\\fl
%\langle\df_{\vec{k}_{1}}(\eta^{*})\df_{\vec{k}_{2}}(\eta^{*})\rangle_{s_{1}s_{2}}=\delta^{(3)}(\vec{k}_{1}+\vec{k}_{2})\frac{H^{4}_{*}}{m_{P}^4}\frac{\ep}{P_{X}c_{s}^{4}}G_{s_{1}s_{2}}(x^{*})F_{s_{1}s_{2}}(k),\\\fl
\langle\df_{\vec{k}_{1}}(\eta^{*})\df_{\vec{k}_{2}}(\eta^{*})\rangle_{s_{1}s_{2}}=\delta^{(3)}(\vec{k}_{1}+\vec{k}_{2})\frac{H^{4}_{*}}{M_{P}^2}\frac{\ep}{P_{X}c_{s}^{3}}G_{s_{1}s_{2}}(x^{*})F_{s_{1}s_{2}}(k),\\\fl
\langle\df_{\vec{k}_{1}}(\eta^{*})\df_{\vec{k}_{2}}(\eta^{*})\rangle_{s_{1}s_{3}}=\delta^{(3)}(\vec{k}_{1}+\vec{k}_{2})\frac{H^{4}_{*}}{M_{P}^2}\frac{\ep}{P_{X}c_{s}^{4}}G_{s_{1}s_{3}}(x^{*})F_{s_{1}s_{3}}(k),\\\fl
%\langle\df_{\vec{k}_{1}}(\eta^{*})\df_{\vec{k}_{2}}(\eta^{*})\rangle_{s_{2}s_{2}}=\delta^{(3)}(\vec{k}_{1}+\vec{k}_{2})\frac{H^{4}_{*}}{m_{P}^4}\frac{\ep^2}{P_{X}c_{s}^{2}}G_{s_{2}s_{2}}(x^{*})F_{s_{2}s_{2}}(k),\\\fl
%\langle\df_{\vec{k}_{1}}(\eta^{*})\df_{\vec{k}_{2}}(\eta^{*})\rangle_{s_{2}s_{3}}=\delta^{(3)}(\vec{k}_{1}+\vec{k}_{2})\frac{H^{4}_{*}}{m_{P}^4}\frac{\ep^2}{P_{X}c_{s}^{3}}G_{s_{2}s_{3}}(x^{*})F_{s_{2}s_{3}}(k),\\\fl
%\langle\df_{\vec{k}_{1}}(\eta^{*})\df_{\vec{k}_{2}}(\eta^{*})\rangle_{s_{2}s_{4}}=\delta^{(3)}(\vec{k}_{1}+\vec{k}_{2})\frac{H^{4}_{*}}{m_{P}^4}\frac{\ep^2}{P_{X}c_{s}^{4}}G_{s_{2}s_{4}}(x^{*})F_{s_{2}s_{4}}(k),\\\fl
\langle\df_{\vec{k}_{1}}(\eta^{*})\df_{\vec{k}_{2}}(\eta^{*})\rangle_{s_{2}s_{2}}=\delta^{(3)}(\vec{k}_{1}+\vec{k}_{2})\frac{H^{4}_{*}}{M_{P}^2}\frac{\ep^2}{P_{X}c_{s}^{4}}G_{s_{2}s_{2}}(x^{*})F_{s_{2}s_{2}}(k),\\\fl
\langle\df_{\vec{k}_{1}}(\eta^{*})\df_{\vec{k}_{2}}(\eta^{*})\rangle_{s_{2}s_{3}}=\delta^{(3)}(\vec{k}_{1}+\vec{k}_{2})\frac{H^{4}_{*}}{M_{P}^2}\frac{\ep^2}{P_{X}c_{s}^{5}}G_{s_{2}s_{3}}(x^{*})F_{s_{2}s_{3}}(k),\\\fl
\langle\df_{\vec{k}_{1}}(\eta^{*})\df_{\vec{k}_{2}}(\eta^{*})\rangle_{s_{3}s_{3}}=\delta^{(3)}(\vec{k}_{1}+\vec{k}_{2})\frac{H^{4}_{*}}{M_{P}^2}\frac{\ep^2}{P_{X}c_{s}^{6}}G_{s_{3}s_{3}}(x^{*})F_{s_{3}s_{3}}(k).
\eea 

\noindent As we can see from the previous equations, tensor loop corrections are subdominant compared to the scalar ones. The details of the computation of the loop diagrams can be found in Appendices C and D. \\

%%%%%%%%%%%%%%%%%%%%%%%%%%%%%%%%%%%%%%%%%%%%%%%%%%%%%%%%%%%%%%%%%%%%%%%%%%%%%%%%%%%%%%%%%%%%%%%%%%%%%%%%%%%%%%%%%%%%%%%%%%%%%%%%%%%%%%%%%%%%%%%%%%%%%%%%%%%%%%%%%%%%%%%%%%%%%%%%%%%%%%%%%%%%%%%%%%%%%%%%%%%%%%%%%%%%%%%%%%%%%%%%%%%%%%%%%%%%%%%%%%%%%%%%%%%%%%%%%%%%%%%%%%%%%%%%%%%%%%%%%%%%%%%%%%%%%%%%%%%%%%%%%%%%%%%%%%%%%%%%%%%%%%%%%%%%%%%%%%%%%%%%%%%%%%%%%%%%%%%%%%%%%%%%%%%%%%%%%%%%%%%%%%%%%%%%%%%%%%%%%%%%%%%%%%%%%%%%%%%%%%%%%%%%%%%%%%%%%%%%%%%%%%%%%%%%%%%%%%%%%%%%%%%%%%%%%%%%%%%%%%%%%%%%%%%%%%%%

\section{Final results}\label{finalresults}

From the expressions of the previous section the leading final result for the one-loop correction to the power spectrum of $\df$ turns out to be
\bea\label{risf1}\fl
\langle\df_{\vec{k}_{1}}(\eta^{*})\df_{\vec{k}_{2}}(\eta^{*})\rangle_{1loop}&=&(2\pi)^3\delta^{(3)}(\vec{k}_{1}+\vec{k}_{2})\frac{H^4}{8\pi^2\ep c_{s}^{6} P_{X}M_{P}^{2}}\Bigg[\left(\frac{8}{9}G_{aa}-\frac{8}{6}G_{ab}+\frac{1}{2}G_{bb}\right)\frac{\ln\left(\Lambda/H_{*}\right)}{k^3}\Bigg].\nonumber\\
\eea
We are now ready to switch from the $\df$ to the $\zeta$ power spectrum. The first line of Eq.~(\ref{QQQ}) (i.e. the one-loop ``quantum'' correction) becomes
\bea\label{fr4}\fl
\langle \zeta_{\vec{k_{1}}}(t)\zeta_{\vec{k_{2}}}(t) \rangle&\supset&(2 \pi)^{3}\delta^{(3)}(\vec{k_{1}}+\vec{k_{2}})\left( N^{(1)} \right)^{2}\left[P_{{\rm tree}}(k)
+P_{{\rm one-loop}}(k)\right] \\\fl&=&(2\pi)^3\delta^{(3)}(\vec{k_{1}}+\vec{k_{2}})\mathcal{P_{\zeta}}\frac{2\pi^2}{k^3}
\left[1+\frac{2 \mathcal{P_{\zeta}}}{c_{s}^{4}}\left(\frac{8}{9}G_{aa}-\frac{8}{6}G_{ab}+\frac{1}{2}G_{bb}\right)\ln\left(\Lambda/H_{*}\right)\right]\, , \nonumber 
\eea
where Eq.~({\ref{tree}) and the equation $\left(N^{(1)}\right)^2= (M_{P}^{-2})(P_{X}/2\ep)$ were employed ($\mathcal{P_{\zeta}}\equiv H^2/(8 \pi^2 M_{P}^{2}\ep c_{s})$, reintroducing the Planck mass). \\

%It is worth noticing that quantum corrections from diagrams with more than one loop do not add any stricter bound than the previous one on the sound speed, as we can see from the following equation ... \textbf{vedi se aggiungere o no le equazioni relative a n loops}  \\

\noindent The final result for $P_{\zeta}$ to one loop is obtained by summing up the contribution we just computed to the ``classical'' loops contributions indicated in the second, third and fourth lines of Eq.~(\ref{QQQ}), i.e.
\bea\label{fr35}\fl
\langle \zeta_{\vec{k_{1}}}(t)\zeta_{\vec{k_{2}}}(t) \rangle&\supset&(2 \pi)^{3}\delta^{(3)}(\vec{k_{1}}+\vec{k_{2}})\Big[N^{(1)}N^{(2)}\int \frac{d^{3}q}{(2 \pi)^{3}} B_{\phi}(k_1,q,|\vec{k}_1-\vec{q}|)\nonumber\\
&+&\frac{1}{2}\left( N^{(2)} \right)^{2}\int \frac{d^{3}q}{(2 \pi)^{3}} P_{{\rm tree}}(q)P_{{\rm tree}}(|\vec{k}_1-\vec{q}|)
\nonumber\\\fl
&+&N^{(1)}N^{(3)}P_{{\rm tree}}(k)\int \frac{d^{3}q}{(2 \pi)^{3}} P_{{\rm tree}}(q)\Big].
\eea
The derivatives of $N$ are given by
\bea
N^{(1)}\simeq \frac{1}{M_{P}}\sqrt{\frac{P_{X}}{\ep}},\quad\quad N^{(2)}\simeq \frac{P_{X}}{M_{P}^{2}},\quad\quad N^{(3)}\simeq\frac{\sqrt{\ep P_{X}^3}}{M_{P}^{3}},
\eea
where we used the definition of the number of e-foldings $N=\int H dt$ together with the results of Sec.~\ref{beom}. Considering Eq.~(\ref{tree}) and computing the order of magnitude of $B_{\phi}$ \footnote{We compute the order of magnitude of tree level diagrams arising from $V_{a}$ and $V_{b}$ which have been found to be the leading order three scalar vertices.}, Eq.~(\ref{fr35}) can be put in the form
\bea\label{fr36}\fl
\langle \zeta_{\vec{k_{1}}}(t)\zeta_{\vec{k_{2}}}(t) \rangle&\supset&(2 \pi)^{3}\delta^{(3)}(\vec{k_{1}}+\vec{k_{2}})\mathcal{P_{\zeta}}^2\left(\frac{2\pi^2}{k^3}\right)\left(\frac{4\ep}{c_{s}^{2}}\right)lnkL\, .
\eea
In deriving Eq.~(\ref{fr36}) the integrals in~(\ref{fr35}) were solved following the method proposed in \cite{Boubekeur:2005fj} (the result in (\ref{fr36}) is exact up to an $\mathcal{O}(1)$ numerical coefficient); $L^{-1}$ is an infrared cutoff that is generally chosen to be comparable to the size of the present cosmological horizon (see Sec.~4 of \cite{seery2} for a complete discussion and for more references on this).\\
By looking at the power spectrum of $\zeta$ with quantum (\ref{fr4}) and classical (\ref{fr36}) one-loop corrections, we can conclude that, logarithmic factors apart, 
quantum corrections are larger than the classical ones.\\ 

Using Eq.~(\ref{fr4}), it is possible to derive some theoretical bounds on the speed of sound, by requesting the one-loop correction not to overcome the tree-level contribution 
for the standard perturbative approach to hold
\bea\label{fr333}
\frac{2\mathcal{P_{\zeta}}}{c_{s}^{4}}\leq 1\quad\quad\quad\quad\Rightarrow\quad\quad\quad\quad c_{s}^{4}\geq 2\mathcal{P_{\zeta}}=\frac{H^2}{4\pi^2\ep c_{s}M_{P}^2}\, ,
\eea
where we have neglected a logarithmic term together with other $\mathcal{O}(1)$ numerical coefficients from the $G_{ij}$ factors.
\footnote{Using Eqs.~(\ref{fr1}) through (\ref{fr3}), the sum appearing in Eq.~(\ref{fr4}) turns out out to be $\left(8G_{aa}/9-8G_{ab}/6+G_{bb}/2\right)
\simeq 1.16$ for $x_*\simeq 1$.} 
On the other hand 
CMB observations allow a measurement of the amplitude of the primordial power spectrum, giving ${\cal P}_{\zeta} \simeq 2.4 \times 10^{-9}$, so that the corresponding bound 
for the sound speed is given by 
\begin{equation}
c_s \gtrsim 0.9 \times 10^{-2}\, .
\end{equation}
It is interesting to note that such a bound is very close to the observational constraint of Ref.~\cite{SSZ},  
$c_s \geq 1.1 \times 10^{-2}$ ($95\%$ C.L.), derived through a detailed analysis of the CMB bispectrum using WMAP 5-year data.
\footnote{Having shown that the the tensor loop corrections are negligible compared to the leading-order scalar contributions, 
the bound we derived for $c_{s}$ is comparable in order of magnitude with one of the bounds obtained in \cite{Leblond:2008gg} (see Eq.~(3.9) therein). See also 
the different approach of Ref.~\cite{Cheung}.} 
The more and more refined techniques of analysis for the CMB bispectrum used  
at present and the high precision data available, clearly show that a theoretical prediction at the same level of precision is indeed required.

\section{Conclusions}
In this paper we have performed a thorough computation of the power spectrum of curvature perturbations $\zeta$ 
arising in general single field models of inflation with a non-standard 
kinetic term. In the spirit of Ref.~\cite{Leblond:2008gg}, the main motivation comes from the fact 
that in these models large non-Gaussianities can be produced when the sound speed $c_s$ is small, 
with an amplitude of the three-point correlation function 
$\langle \zeta \zeta \zeta\rangle$ given by the non-linearity parameter 
$f_{\rm NL}\sim c^{-2}_s$~\cite{Chen:2006nt}. This suggests that, when computing the two point function $\langle \zeta \zeta \rangle$ up to one-loop,  
strong corrections could arise as well, and invalidate the usual perturbation theory approach. A rigorous computation is required to see 
the precise conditions for which the one-loop corrections are under control. 
In doing so we have obtained various results. We have provided the general action at fourth-order 
(see Appendix B), and we have included the tensor fluctuations as well, showing explicitly that the loop-corrections involving tensor mode fluctuations are negligible 
with respect to the pure scalar contributions. In deriving the leading order corrections at one-loop to the power spectrum of the curvature perturbation, we have 
expressed the magnitude of the various interaction terms appearing in the Lagrangian in terms of slow-roll variation parameters and the sound speed, as summarized in Table I. 
We have applied these results to various cases, which include also DBI models. 
For the usual perturbative approach to remain valid, the one-loop corrrections must be smaller than the tree-level power spectrum and this 
imposes an upper bound on the sound speed, $c_s \gtrsim {\cal O}(1) \times 10^{-2}$.
Interestingly enough such a bound turns out to be very close to the observational bound derived in Ref.~\cite{SSZ}.

%%%%%%%%%%%%%%%%%%%%%%%%%%%%%%%%%%%%%%%%%%%%%%%%%%%%%%%%%%%%%%%%%%%%%%%%%%%%%%%%%%%%%%%%%%%%%%%%%%%%%%%%%%%%%%%%%%%%%
%%%%%%%%%%%%%%%%%%%%%%%%%%%%%%%%%%%%%%%%%%%%%%%%%%%%%%%%%%%%%%%%%%%%%%%%%%%%%%%%%%%%%%%%%%%%%%%%%%%%%%%%%%%%%%%%%%%%%%
%%%%%%%%%%%%%%%%%%%%%%%%%%%%%%%%%%%%%%%%%%%%%%%%%%%%%%%%%%%%%%%%%%%%%%%%%%%%%%%%%%%%%%%%%%%%%%%%%%%%%%%%%%%%%%%%%%%%%
%%%%%%%%%%%%%%%%%%%%%%%%%%%%%%%%%%%%%%%%%%%%%%%%%%%%%%%%%%%%%%%%%%%%%%%%%%%%%%%%%%%%%%%%%%%%%%%%%%%%%%%%%%%%%%%%%%%%%
%%%%%%%%%%%%%%%%%%%%%%%%%%%%%%%%%%%%%%%%%%%%%%%%%%%%%%%%%%%%%%%%%%%%%%%%%%%%%%%%%%%%%%%%%%%%%%%%%%%%%%%%%%%%%%%%%%%%%
%%%%%%%%%%%%%%%%%%%%%%%%%%%%%%%%%%%%%%%%%%%%%%%%%%%%%%%%%%%%%%%%%%%%%%%%%%%%%%%%%%%%%%%%%%%%%%%%%%%%%%%%%%%%%%%%%%%%%
%%%%%%%%%%%%%%%%%%%%%%%%%%%%%%%%%%%%%%%%%%%%%%%%%%%%%%%%%%%%%%%%%%%%%%%%%%%%%%%%%%%%%%%%%%%%%%%%%%%%%%%%%%%%%%%%%%%%

\section*{Acknowledgments}
We are happy to thank Peter\,Adshead, Xingang\,Chen, Richard\,Easther, Eugene\,A.\,Lim, Sabino\,Matarrese and Massimo\,Pietroni for useful discussions. We especially thank David\,Seery and Sarah\,Shandera for important correspondence. This research has been partially supported by the ASI
Contract No. I/016/07/0 COFIS and by the ASI/INAF Agreement I/072/09/0 for
the Planck LFI Activity of Phase E2. AV is supported by the DOE at Fermilab.

\setcounter{equation}{0}
\def\theequation{A\arabic{equation}}
\section*{Appendix A. Solutions for the Hamiltonian and momentum constraints}

In the spatially flat gauge and using the ADM formalism, the perturbed metric is
\be
ds^2=-N^2dt^2+h_{ij}(dx^{i}+N^{i}dt)(dx^{j}+N^{j}dt)
\ee
where
\be
h_{ij}=a^{2}(t)(e^{\ga})_{ij}
\ee
with $\ga_{ij}$ traceless and divergenceless. The lapse and the shift function can be expanded as follows
\bea
N=1+\al \label{exp1},\\
N_{j}=\p_{j}\th+\b_{j} \label{exp2},
\eea
where $\b_{j}$ is divergenceless and $\al$ and $\th$ are scalar fluctuations. Before deriving the Hamiltonian and the momentum constraints, we also need to perturbatively expand $P(X,\phi)$. Let us begin with the expansion of $X$ 
\bea
X&=&-g^{\mu\nu}\p_{\mu}\phi\p_{\nu}\phi=-\frac{1}{2}\left[g^{00}\dot{{\phi}^{2}}+2g^{0i}\p_{i}\phi \dot{\phi}+g^{ij}\p_{i}\phi\p_{j}\phi\right]=\nonumber\\
&=&-\frac{1}{2}\Big[-N^{-2}{\left(\dot{\phi}+\dot{\df}\right)}^{2}+2N^{-2}N^{i}\p_{i}\df\left(\dot{\phi}+\dot{\df}\right)\nonumber\\
&+&\left(h^{ij}-\frac{N^{i}N^{j}}{N^{2}}\right)\p_{i}\df\p_{j}\df\Big]=X_{0}+\Delta X,
\eea
where $N^{i}\equiv h^{ij}N_{j}$, $X_{0}$ is the zeroth order part, i.e. $X_{0}={\dot{\phi}}^{2}/2$ and $\Delta X$ is the perturbation to the desired order. Notice that  $\phi(t,\vec{x})=\phi_{0}(t)+\df(t,\vec{x})$, but for simplicity we will suppress the subscript '${0}$' in the background value of the field (the order at which $\phi$ is to be taken will be clear based on the particular context). \\
The expressions for the perturbations $\De X_{i}$ have been completed as follows: the quantity $X_{0}$ has been factorized
\bea\fl
\Delta X_{1}&=&2X_{0}\left[\frac{\dot{\df}}{\dot{\phi}}-\alo\right] \\\fl
\Delta X_{2}&=&X_{0}\Big[{\left(\frac{\ddf}{\dP}\right)}^{2}-4\alo\frac{\ddf}{\dP}-2\alt+3{\alo}^{2}-2N^{i1}\p_{j}\frac{\ddf}{\dP}\Big]-\frac{1}{a^{2}{\dP}^{2}}\p_{i}\df\p_{i}\df\\\fl
\Delta X_{3}&=&X_{0}\Big[-2\alo{\left(\frac{\ddf}{\dP}\right)}^{2}-4\alt\frac{\ddf}{\dP}+6{\alo}^{2}\frac{\ddf}{\dP}+6\alo\alt-4{\alo}^{3}-2N^{i2}\p_{i}\frac{\df}{\dP}\nonumber\\\fl
&+&4N^{i1}\alo\p_{i}\frac{\df}{\dP}-2N^{i1}\frac{\ddf}{\dP}\p_{i}\frac{\df}{\dP}+a^{-2}\delta^{il}\delta^{jm}\ga_{lm}\p_{i}\frac{\df}{\dP}\p_{j}\frac{\df}{\dP}\Big]\\ \fl
\Delta X_{4}&=&X_{0}\Big[-2\alt{\left(\frac{\ddf}{\dP}\right)}^{2}+3{\alo}^{2}{\left(\frac{\ddf}{\dP}\right)}^{2}+3{\alt}^{2}+12\alo\alt\frac{\ddf}{\dP}-8{\alo}^{3}\frac{\ddf}{\dP}-12{\alo}^{2}\alt+5{\alo}^{4}\nonumber\\\fl
&-&2a^{-2}\Big(\frac{1}{2}\delta^{rj}\delta^{il}\delta^{st}\ga_{ls}\ga_{tr}N_{j}^{1}\p_{i}\frac{\df}{\dP}+2\alo\de^{il}\de^{jm}\ga_{lm}N^{1}_{j}\p_{i}\frac{\df}{\dP}-2\alt\delta^{ij}N^{1}_{j}\p_{i}\frac{\df}{\dP}\nonumber\\\fl
&+&3{\alo}^{2}\de^{ij}N^{1}_{j}\p_{i}\frac{\df}{\dP}-\delta^{il}\delta^{jm}\ga_{lm}N^{2}_{j}\p_{i}\frac{\df}{\dP}\nonumber\\\fl
&-&2\alo\delta^{ij}N^{2}_{j}\p_{i}\frac{\df}{\dP}-\delta^{il}\delta^{jm}\ga_{lm}N^{1}_{j}\p_{i}\frac{\df}{\dP}\frac{\ddf}{\dP}-2\delta^{ij}\alo\frac{\ddf}{\dP}N^{1}_{j}
\p_{i}\frac{\ddf}{\dP}+\delta^{ij}N^{2}_{j}\frac{\ddf}{\dP}\p_{i}\frac{\df}{\dP}\Big)\nonumber\\\fl
&-&\frac{1}{2a^{2}}\delta^{rj}\delta^{il}\delta^{st}\ga_{ls}\ga_{tr}\p_{i}\frac{\df}{\dP}\p_{j}\frac{\df}{\dP}+\frac{1}{a^{4}}\delta^{ij}\delta^{lm}N_{i}N_{l}\p_{j}\frac{\df}{\dP}\p_{m}\frac{\df}{\dP}\Big].
\eea
Up to fourth order we have
\bea\fl
P(X,\phi)&=&P+P_{X}\De X+P_{\phi}\df+\frac{1}{2!}P_{XX}{\De X}^{2}+\frac{1}{2!}P_{\phi\phi}{\df}^{2}+P_{X\phi}\De X \df\nonumber\\\fl&+&
\frac{1}{3!}P_{XXX}{\De X}^{3}+\frac{1}{3!}P_{\phi\phi\phi}{\df}^{3}+\frac{1}{2!}P_{XX\phi}{\De X}^{2}\df+\frac{1}{2!}P_{X\phi\phi}\De X {\df}^{2}\nonumber\\\fl&+&
\frac{1}{4!}P_{XXXX}{\De X}^{4}+\frac{1}{4!}P_{\phi\phi\phi\phi}{\df}^{4}+\frac{1}{3!}P_{XXX\phi}{\De X}^{3}+
\frac{1}{3!}P_{X\phi\phi\phi}\De X{\df}^{3}\nonumber\\\fl&+&
\frac{1}{4}P_{XX\phi\phi}{\De X}^{2}{\df}^{2}\nonumber\\\fl&=&
P+P_{X}\left(\De X_{1}+\De X_{2}+\De X_{3}+\De X_{4}\right)+P_{\phi}\df\nonumber\\\fl&+&\frac{1}{2!}P_{XX}\left({\De X_{1}}^{2}+{\De X_{2}}^{2}+
2\De X_{1}\De X_{2}+2\De X_{1}\De X_{3}\right)\nonumber\\\fl&+&\frac{1}{2!}P_{\phi\phi}{\df}^{2}+P_{X\phi}\left(\De X_{1}+\De X_{2}+\De X_{3}\right)\df\nonumber\\\fl&+&
\frac{1}{3!}P_{XXX}\left({\De X_{1}}^{3}+3{\De X_{1}}^{2}\De X_{2}\right)+\frac{1}{3!}P_{\phi\phi\phi}{\df}^{3}\nonumber\\\fl&+&
\frac{1}{2!}P_{XX\phi}\left({\De X_{1}}^{2}+2\De X_{1}\De X_{2}\right)\df+\frac{1}{2!}P_{X\phi\phi}\left(\De X_{1}+\De X_{2}
\right) {\df}^{2}\nonumber\\\fl&+&
\frac{1}{4!}P_{XXXX}{\De X_{1}}^{4}+\frac{1}{4!}P_{\phi\phi\phi\phi}{\df}^{4}+\frac{1}{3!}P_{XXX\phi}{\De X_{1}}^{3}\df\nonumber\\\fl&+&
\frac{1}{3!}P_{X\phi\phi\phi}\De X_{1}{\df}^{3}+
\frac{1}{4}P_{XX\phi\phi}{\De X_{1}}^{2}{\df}^{2},
\eea
where $P_{X}=\p_{X}P$, $P_{\phi}=\p_{\phi}P$ and so on for higher order derivatives (we have for simplicity suppressed the subscript '$0$' on the right hand side in $P$ and its derivatives).\\
The kinetic part of the action is explicitly given to fourth order by
\be
2NP=2\left(1+\alo+\alt\right)P=2P^{(4)}+2\alo P^{(3)}+2\alt P^{(2)},
\ee
where
\bea
P^{(2)}&\equiv& P_{X}\De X_{2}+\frac{1}{2!}P_{XX}{\De X_{1}}^{2}+\frac{1}{2!}P_{\phi\phi}{\df}^{2}+P_{X\phi}\De X_{1}\df,\\
P^{(3)} &\equiv& P_{X}\De X_{3}+P_{XX}\De X_{1}\De X_{2}+P_{X\phi}\De X_{2}\df\nonumber\\&+&
\frac{1}{3!}P_{XXX}{\De X_{1}}^{3}+\frac{1}{3!}P_{\phi\phi\phi}{\df}^{3}+\frac{1}{2!}P_{XX\phi}{\De X_{1}}^{2}\df\nonumber\\
&+&\frac{1}{2!}P_{X\phi\phi}\De X_{1}{\df}^{2}+\frac{1}{3!}P_{XXX\phi}{\De X_{1}}^{3}, \\
P^{(4)} &\equiv& P_{X}\De X_{4}\nonumber\\&+&
\frac{1}{2!}P_{XX}\left({\De X_{2}}^{2}+2\De X_{1}\De X_{3}\right)+P_{X\phi}\De X_{3}\df+\frac{1}{2!}P_{XXX}{}\De X_{1}^{2}\De X_{2}\nonumber\\&+&
P_{XX\phi}\De X_{1}\De X_{2}\df+\frac{1}{2!}P_{X\phi\phi}\De X_{2}{\df}^{2}+\frac{1}{4!}P_{XXXX}{\De X_{1}}^{4}\nonumber\\&+&
\frac{1}{4!}P_{\phi\phi\phi\phi}{\df}^{4}+\frac{1}{3!}P_{X\phi\phi\phi}\De X_{1}{\df}^{3}+\frac{1}{4}P_{XX\phi\phi}{\De X_{1}}^{2}{\df}^{2}\nonumber\\
&+&\frac{1}{3!}P_{XXX\phi}{\De X_{1}}^{3}\df.
\eea
The next step consists in writing and solving momentum and hamiltonian contraints in order to integrate out the lapse and the shift functions $N$ and $N_{i}$. This was done in \cite{Dimastrogiovanni:2008af}, where solutions for momentum and hamiltonian constraint were provided for a general Lagrangian. \\
The constraint equations are
\be
\bt_{i}\left[N^{-1}\left(E^{i}_{j}-\de^{i}_{j} E\right)\right]=N^{-1}P_{,X}\left[\dot{\phi}-N^{l}\p_{l}\phi\right]\p_{j}\phi,
\ee
\be
R^{(3)}+2P-4P_{,X}X-N^{-2}\left(E_{ij}E^{ij}-E^{2}\right)-2P_{,X}h^{ij}\p_{i}\phi\p_{j}\phi=0.
\ee
The action to a given order $n$ only requires the constraint equations to be solved up to order $n-2$. Therefore we will solve the constraints to second order in the metric and scalar field fluctuations. Let's employ the expansions
\bea
\al=\alo+\alt, \label{exp3} \\
\b_{i}=\b_{1i}+\b_{2i}, \label{exp4}\\
\th=\tho+\tht \label{exp5}.
\eea
The momentum constraint to first order is as follows
\be
2H\p_{j}\alo-\frac{1}{2a^{2}}\p_{i}\p^{i}N_{j}-\frac{1}{2a^{2}}\p_{i}\p_{j}N^{i}+a^{-2}\p_{j}\left(\p^{q}N_{q}\right)=P_{,X}\dot{\phi}\p_{j}\df,
\ee
where $\p^{i} \equiv \delta^{ij}\p_{j}$ and $N^{i} \equiv {\de}^{ij}N_{j}$. \\
This can be solved to get $\alo$: deriving both sides by $\p^{j}$ and using the divergenceless condition for $\b$ the result is
\be
\alo=\frac{P_{,X}\dP\df}{2H}.
\ee
Going one step behind and using the solution found for $\alo$, we find $\p^{i}\p_{i}\b_{1j}$. The solution $\b_{1j}=0$ can be chosen. From now on we will define $\b_{i} \equiv \b_{2i}$ for simplicity.\\
The momentum constraint to second order is
\bea
2H\p_{j}\alt-\frac{1}{2a^{2}}\p^{2}\b_{j}-4H\alo\p_{j}\alo-\frac{1}{a^{2}}\p_{i}\alo\left(\de^{i}_{j}\p^{2}\tho-\de^{il}\p_{l}\p_{j}\tho\right)\\
-\frac{1}{2}\dot{\ga_{lj}}\de^{li}\p_{i}\alo-\frac{1}{2a^{2}}\p_{j}\ga_{bq}\de^{bs}\de^{qr}\p_{s}\p_{r}\tho+\frac{1}{2a^{2}}\p^{2}\ga_{jk}\de^{kl}\p_{l}\tho-\frac{1}{4}\ga_{il}\de^{ik}\de^{lp}\p_{k}\dot{\ga_{jp}}\nonumber\\
+\frac{1}{4}\dot{\ga_{ik}}\de^{il}\de^{kr}\p_{l}\ga_{rj}-\ddf-\p_{j}\df-\alo\dP\p_{j}\df=0\nonumber
\eea
where $\p^{2}\equiv \de^{ij}\p_{i}\p_{j}$.\\
The solutions are
\bea
\alt&=&\frac{\alo^2}{2}+\frac{1}{2Ha^2}\p^{-2}\left[\p^{2}\alo\p^{2}\tho-\p_{i}\p_{j}\alo\de^{lm}\p_{l}\p_{m}\tho\right]+\frac{P_{,X}}{2H}\p^{-2}\Sigma\nonumber\\
&+&\frac{1}{4H}\p^{-2}\left[\dot{\ga_{ij}}\de^{lm}\p_{l}\p_{m}\alo+\frac{1}{a^{2}}\p_{j}\ga_{bq}\de^{js}\de^{br}\de^{qt}\p_{s}\p_{r}\p_{t}\tho\right]\nonumber\\
&+&\frac{XP_{,XX}}{H}\p^{-2}\left[\p^{2}\df\ddf+\de^{ij}\p_{i}\df\p_{j}\ddf-\dP\left(\de^{ij}\p_{i}\alo\p_{j}\df+\alo\p^{2}\df\right)\right]\nonumber\\
&+&\frac{P_{,X\ph}\dP}{2H}\p^{-2}\left[\de^{ij}\p_{i}\df\p_{j}\df+\df\p^{2}\df\right]
\eea
and
\bea
\frac{\b_{j}}{2a^{2}}&=& \p^{-2}\big[2H\p_{j}\alt-4H\alo\p_{j}\alo-\frac{1}{a^{2}}\p_{i}\alo\left(\de^{i}_{j}\p^{2}\tho-\de^{il}\p_{l}\p_{j}\tho\right)\big]\nonumber\\
&+&\p^{-2}\big[-\frac{1}{2}\dot{\ga_{ij}}\de^{il}\p_{l}\alo-\frac{1}{2a^2}\p_{j}\ga_{bq}\de^{br}\de^{qs}\p_{r}\p_{s}\tho+\frac{1}{2a^2}\p^2\ga_{jk}\de^{kl}\p_{l}\tho\nonumber\\&-&\frac{1}{4}\ga_{il}\de^{ir}\de^{ls}\p_{r}\dot{\ga_{js}}\big]+\p^{-2}\big[-P_{,X\phi}\df\dP\p_{j}\df+\dP P_{,X}\alo\p_{j}\df\big]\\
&+&\p^{-2}\big[\frac{1}{4}\dot{\ga_{ik}}\de^{ir}\de^{ks}\p_{r}\ga_{sj}-P_{,X}\p_{j}\df\ddf-2X\dot{\phi}P_{,XX}\p_{j}\df\left(\frac{\ddf}{\dP}-\alo\right)\big],\nonumber
\eea
for the vector mode.\\
Let's move now to the hamiltonian constraint. To first order it is solved by
\bea
\frac{4H}{a^2}\p^{2}\tho&=&-4XP_{,X}\left(\frac{\ddf}{\dP}-\alo\right)\nonumber\\
&+&2P_{,\ph}\df-8P_{,XX}X^2\left(\frac{\ddf}{\dP}-\alo\right)-4XP_{,X\ph}\df-12H^2\alo.
\eea
To second order
\bea\fl
-\frac{4H}{a^2}\p^{2}\tht&=&
(-2\alo)\big[4XP_{,X}\frac{\ddf}{\dP}+20P_{,XX}X^{2}\frac{\ddf}{\dP}+2XP_{,X\phi}\df+8P_{,XXX}X^{3}\frac{\ddf}{\dP}\nonumber\\\fl&+&4P_{,XX\phi}X^{2}\df+\frac{4}{a^{2}}\p^{2}\tho\big]-\frac{4X\left(P_{,X}+2XP_{,XX}\right)}{a^{2}\dP}\de^{ij}\p_{i}\tho\p_{j}\df\nonumber\\\fl&-&\frac{1}{a^{4}}\left(\p^{2}\tho\right)^{2}+\frac{1}{a^{2}}\left[-\dot{\ga_{iq}}\de^{ir}\de^{qs}\p_{s}\p_{r}\tho+\de^{il}\de^{jm}\p_{i}\p_{j}\tho\p_{l}\p_{m}\tho\right]\nonumber\\\fl
&\times&\left(-6H^{2}+2XP_{,X}+4X^{2}P_{,XX}\right)\left[3\alo^{2}-2\alt\right]+4\alo^{2}\left(3X^{2}P_{,XX}+2X^{3}P_{,XXX}\right)\nonumber\\\fl
&+&\frac{\ddf^{2}}{\dP^{2}}\left[2XP_{,X}+16X^{2}P_{,XX}+8X^{3}P_{,XXX}\right]+\frac{\ddf\df}{\df}\left[4XP_{,X\phi}+8XP_{,XX\phi}\right]\nonumber\\\fl
&-&\de^{ij}\p_{i}\df\p_{j}\df\left[\frac{4X^{2}P_{,XX}}{a^{4}\dP^{2}}+\frac{2P_{,X}}{a^{2}}\right]
+\frac{1}{4}\left[\dot{\ga_{lj}}\de^{lr}\de^{jr}\dot{\ga_{rs}}+\frac{1}{a^{2}}\p_{a}\ga_{iq}\de^{ar}\de^{is}\de^{qt}\p_{r}\ga_{st}\right].\nonumber\\
\eea
We are now ready to expand the action (\ref{action-general}) up to fourth order in perturbation theory. The result is provided in Appendix B.

%%%%%%%%%%%%%%%%%%%%%%%%%%%%%%%%%%%%%%%%%%%%%%%%%%%%%%%%%%%%%%%%%%%%%%%%%%%%%%%%%%%%%%%%%%%%%%%%%%%%%%%%%%%%%%%%%%%%%
%%%%%%%%%%%%%%%%%%%%%%%%%%%%%%%%%%%%%%%%%%%%%%%%%%%%%%%%%%%%%%%%%%%%%%%%%%%%%%%%%%%%%%%%%%%%%%%%%%%%%%%%%%%%%%%%%%%%%
%%%%%%%%%%%%%%%%%%%%%%%%%%%%%%%%%%%%%%%%%%%%%%%%%%%%%%%%%%%%%%%%%%%%%%%%%%%%%%%%%%%%%%%%%%%%%%%%%%%%%%%%%%%%%%%%%%%%%
%%%%%%%%%%%%%%%%%%%%%%%%%%%%%%%%%%%%%%%%%%%%%%%%%%%%%%%%%%%%%%%%%%%%%%%%%%%%%%%%%%%%%%%%%%%%%%%%%%%%%%%%%%%%%%%%%%%%%
%%%%%%%%%%%%%%%%%%%%%%%%%%%%%%%%%%%%%%%%%%%%%%%%%%%%%%%%%%%%%%%%%%%%%%%%%%%%%%%%%%%%%%%%%%%%%%%%%%%%%%%%%%%%%%%%%%%%%
%%%%%%%%%%%%%%%%%%%%%%%%%%%%%%%%%%%%%%%%%%%%%%%%%%%%%%%%%%%%%%%%%%%%%%%%%%%%%%%%%%%%%%%%%%%%%%%%%%%%%%%%%%%%%%%%%%%%%
%%%%%%%%%%%%%%%%%%%%%%%%%%%%%%%%%%%%%%%%%%%%%%%%%%%%%%%%%%%%%%%%%%%%%%%%%%%%%%%%%%%%%%%%%%%%%%%%%%%%%%%%%%%%%%%%%%%%%
%%%%%%%%%%%%%%%%%%%%%%%%%%%%%%%%%%%%%%%%%%%%%%%%%%%%%%%%%%%%%%%%%%%%%%%%%%%%%%%%%%%%%%%%%%%%%%%%%%%%%%%%%%%%%%%%%%%%%
%%%%%%%%%%%%%%%%%%%%%%%%%%%%%%%%%%%%%%%%%%%%%%%%%%%%%%%%%%%%%%%%%%%%%%%%%%%%%%%%%%%%%%%%%%%%%%%%%%%%%%%%%%%%%%%%%%%%%%%%%%%%%%%%%%%%%%%%%%%%%%%%%%%%%%%%%%%%%%%%%%%%%%%%%%%%%%%%%%%%%%%%%%%%%%%%%%%%%%%%%%%%%%%%%%%%%%%%%%%%%%%%%%%%%%%%%%%%
%%%%%%%%%%%%%%%%%%%%%%%%%%%%%%%%%%%%%%%%%%%%%%%%%%%%%%%%%%%%%%%%%%%%%%%%%%%%%%%%%%%%%%%%%%%%%%%%%%%%%%%%%%%%%%%%%%%%%

\setcounter{equation}{0}
\def\theequation{B\arabic{equation}}
\section*{Appendix B. Complete expressions for the action up to fourth order in perturbation theory}\label{action}

The action to third order is given by 
\bea
S^{(3)}=\frac{1}{2}\int d^{3}x d \eta a^{4}\Big[\textit{L}_{\phi}+\textit{L}_{\ga}+\textit{L}_{int}\Big],
\eea
where
\bea\fl
\textit{L}_{\phi}&=&P_{X}\Big[-\alo\ddf^2+2\fd\alo^2\ddf-\fd^2\alo^3\Big]+\frac{P_{,X}}{a^2}\Big[\alo{\Big(\p_{i}\df\Big)}^{2}-2\ddf\p_{i}\df\p_{i}\tho\Big]\nonumber\\\fl&+&P_{,XX}\Big[\fd\ddf^3-4\fd^2\alo\ddf^2+5\fd^5\alo^2\ddf-2\fd^4\alo^3\Big]+\frac{P_{,XX}}{a^2}\Big[-\fd\ddf{\Big(\p_{i}\df\Big)}^{2}\nonumber\\\fl&+&\fd^2\alo{\Big(\p_{i}\df\Big)}^{2}-2\fd^2\ddf\p_{i}\df\p_{i}\tho+2\fd^3\alo\p_{i}\df\p_{i}\tho\Big]+P_{,XXX}\Big[\frac{1}{3}\fd^3\ddf^3-\fd^4\alo\ddf^2\nonumber\\\fl&+&\fd^5\alo^2\ddf-\frac{1}{3}\fd^6\alo^3\Big]+P_{,X\f}\alo\df^2+\frac{1}{3}P_{,\f\f\f}\df^3+P_{,X\f}\Big[\df\ddf^2-2\fd\alo\df\ddf+\fd^2\alo^2\df\Big]\nonumber\\\fl&+&\frac{P_{,X\f}}{a^2}\Big[\df{\Big(\p_{i}\df\Big)}^{2}-2\fd\df\p_{i}\df\p_{i}\tho\Big]+P_{X\f\f}\Big[\fd\df^2\ddf-\fd^2\alo\df^2\Big]\nonumber\\\fl&+&P_{,XX\f}\Big[\fd^2\df\ddf^2-2\fd^3\alo\df\ddf+\fd^4\alo^2\df\Big]+6H^2\alo^3+\frac{4H}{a^2}\alo^2{\Big(\p_{i}\tho\Big)}^{2}+\frac{1}{a^4}\alo{\Big(\p^{2}\tho\Big)}^{2}\nonumber\\\fl&-&\frac{1}{a^4}\alo{\Big(\p_{i}\p_{j}\tho\Big)}^{2},
\\\fl
\textit{L}_{\ga}&=& H\ga_{ik}\dot{\ga}_{kj}\ga_{ji},\\\fl
\textit{L}_{int}&=& -\frac{1}{4}\alo\dot{\ga}_{ij}\dot{\ga}_{ij}-\frac{1}{2a^2}\dot{\ga}_{ij}\p_{i}\ga_{jk}\p_{k}\tho-\frac{1}{4a^2}\alo\p_{i}\ga_{jk}\p_{i}\ga_{jk}+\frac{4H}{a^2}\alo\ga_{ij}\p_{i}\p_{j}\tho\nonumber\\\fl&+&\frac{2H}{a^2}\ga_{ij}\ga_{jk}\p_{k}\p_{i}\tho+\frac{1}{a^2}\alo\dot{\ga_{ij}}\p_{i}\p_{j}\tho+\frac{1}{a^2}\ga_{ij}\dot{\ga_{jk}}\p_{k}\p_{i}\tho+\frac{2H}{a^2}\ga_{ij}\p_{i}\ga_{jk}\p_{k}\tho\nonumber\\\fl&+&\frac{1}{a^2}\dot{\ga_{ij}}\p_{i}\ga_{jk}\p_{k}\tho-\frac{2}{a^4}\ga_{ij}\p_{i}\p_{k}\tho\p_{j}\p_{k}\tho+\frac{2}{a^4}\ga_{ij}\p_{i}\p_{j}\tho\p^{2}\tho+\frac{1}{a^4}\p_{i}\ga_{jk}\p_{i}\tho\p_{j}\p_{k}\tho\nonumber\\\fl&-&\frac{2}{a^4}\p_{i}\tho\p_{j}\ga_{ik}\p_{j}\p_{k}\tho.
\eea\\

\noindent The complete expression for the fourth order action is

\bea\label{gr11}\fl
S^{(4)}=\int d^{3}x d\eta \frac{a^{4}}{2}&\Bigg[&-6H^2\alo^{2}+18 H^2\alo^2\alt-6H^2\alt^2+P_{\f\f}\alt\df^2+\frac{P_{\f\f\f}}{3}\alo\df^3\nonumber\\\fl&+&\frac{P_{\f\f\f\f}}{12}\df^4-H^2\ga^{ik}\ga_{ij}\ga^{jl}\ga_{kl}+\frac{1}{4}\alo\dot{\ga}_{ij}\dot{\ga}^{ij}-\frac{1}{4}\alt\dot{\ga}_{ij}\dot{\ga}^{ij}\nonumber\\\fl&-&H\alo\ga_{i}^{k}\ga^{ij}\dot{\ga}_{jk}-\frac{1}{8}\ga^{ij}\ga^{kl}\dot{\ga}_{ik}\dot{\ga}_{jl}-H\ga_{i}^{k}\ga^{ij}\ga_{j}^{k}\dot{\ga}_{kl}-\frac{1}{8}\ga_{i}^{k}\ga^{ij}\dot{\ga}_{j}^{l}\dot{\ga}_{kl}\nonumber\\\fl&+&P_{XX\f\f}\Bigg(\frac{\dP^2}{2}\df^2\ddf^2-\dP^3\alo\df^3\ddf+\frac{\dP^4}{2}\alo^2\df^2\Bigg)+P_{X\f\f\f}\Bigg(\dP\df^3\ddf-\frac{\dP^2}{3}\alo\df^3\Bigg)\nonumber\\\fl&+&P_{X\f\f\f}\Bigg(\frac{\dP^3}{3}\df\ddf^3-\dP^4\alo\df^2\ddf+\dP^5\alo^2\df\ddf-\frac{\dP^6}{3}\alo^3\df\Bigg)\nonumber\\\fl&+&P_{XXXX}\Bigg(\frac{\dP^4}{12}\ddf^4-\frac{\dP^5}{3}\alo\ddf^3+\frac{\dP^6}{2}\alo^2\ddf^2-\frac{\dP^7}{3}\alo^3\ddf+\frac{\dP^8}{12}\alo^4\Bigg)\nonumber\\\fl&-&\frac{\b^{i}\dot{\ga}^{jk}\p_{i}\ga_{jk}}{2a^2}-\frac{4H\alo^3\p^2\tho}{a^2}+\frac{8H\alo\alt\p^{2}\tho}{a^2}+\frac{\ga^{l}_{j}\ga^{jk}\dot{\ga}_{kl}\p^{2}\tho}{2a^2}-\frac{4H}{a^2}\alo^2\p^{2}\tht\nonumber\\\fl&-&\frac{4H\alt\p^2\tht}{a^2}+\frac{\alo\dot{\ga}^{jk}\p_{i}\ga_{jk}\p^{i}\tho}{2a^2}-\frac{H\ga^{l}_{j}\ga^{jk}\p_{i}\ga_{kl}\p^{i}\tho}{a^2}-\frac{\dot{\ga}^{jk}\p_{i}\ga_{jk}\p^{i}\tht}{2a^2}\nonumber\\\fl&-&\frac{\delta^{ij}\alt\p_{i}\ga^{kl}\p_{j}\ga_{kl}}{4a^2}+\frac{2H\ga^{j}_{i}\ga^{kl}\p^{i}\tho\p_{j}\ga_{kl}}{a^2}+\frac{\ga^{j}_{i}\dot{\ga}^{kl}\p^{i}\tho\p_{j}\ga_{kl}}{2a^2}-\frac{\alo^2\left(\p^2\tho\right)^2}{a^4}\nonumber\\\fl&+&\frac{\alt\left(\p^2\tho\right)^2}{a^4}+\frac{2\alo\p^{2}\tho\p^{2}\tht}{a^4}-\frac{\left(\p^{2}\tht\right)^2}{a^4}+\frac{4H\alo\ga_{ij}\p^{i}\b^{j}}{a^2}+\frac{2H\ga^{k}_{i}\ga_{jk}\p^{j}\b^{i}}{a^2}\nonumber\\\fl&+&\frac{\alo\dot{\ga}_{ij}\p^{i}\b^{j}}{a^2}+\frac{\ga^{k}_{j}\dot{\ga}_{ik}\p^{j}\b^{i}}{2a^2}+\frac{\ga^{k}_{i}\dot{\ga}_{jk}\p^{j}\b^{i}}{2a^2}+\frac{\p_{i}\b_{j}\p^{j}\b^{i}}{2a^4}+\frac{\p_{i}\b_{j}\p^{i}\b^{j}}{2a^4}\nonumber\\\fl&-&\frac{2\alo\p_{i}\p_{j}\tho\p^{i}\b^{j}}{a^4}+\frac{2\p_{i}\p_{j}\tht\p^{i}\b^{j}}{a^4}+\frac{\p_{i}\ga^{kl}\p^{i}\tho\p_{j}\ga_{kl}\p^{j}\tho}{4a^4}+P_{X\f\f}\Bigg(\frac{\df^2\ddf^2}{2}\nonumber\\\fl&-&\dP\alo\df^2\ddf+\frac{\dP^2\alo^2\df^2}{2}-\dP^2\alt\df^2-\frac{\delta_{ij}\df^2\p_{i}\df\p_{j}\df}{2a^2}-\frac{\dP\df^2\delta_{ij}\p^{i}\df\p^{j}\tho}{a^2}
\Bigg)\nonumber\\\fl&+&P_{XX\f}\Bigg(\dP\df\ddf^3-4\dP^2\alo\df\ddf+5\dP^3\alo^2\df\ddf-2\dP^3\alt\df\ddf-2\dP^4\alo^3\df\nonumber\\\fl&+&2\dP^4\alo\alt\df-\frac{\dP\df\ddf\delta_{ij}\p^{i}\df\p^{j}\df}{a^2}+\frac{\dP^2\alo\df\delta_{ij}\p^{i}\df\p^{j}\df}{a^2}\nonumber\\\fl&-&\frac{2\dP^2\df\ddf\delta_{ij}\p^{i}\df\p^{j}\tho}{a^2}+\frac{2\dP^3\df\delta_{ij}\p^{i}\df\p^{j}\tho}{a^2}\Bigg)+P_{XXX}\Bigg(\frac{\dP^2\ddf^4}{2}-\frac{8\dP^3\alo\ddf^3}{3}\nonumber\\\fl&+&5\dP^4\alo^2\ddf^2-\dP^4\alt\ddf^2-4\dP^5\alo^3\ddf+\dP^5\alo\alt\ddf+\frac{7\dP^6\alo^4}{6}-\dP^6\alo^2\alt\nonumber\\\fl&-&\frac{\dP^2\ddf^2\delta_{ij}\p^{i}\df\p^{j}\df}{2a^2}+\frac{\dP^3\alo\delta_{ij}\p^{i}\df\p^{j}\df}{a^2}+\frac{\dP^4\alo^2\delta_{ij}\p^{i}\df\p^{j}\df}{2a^2}\nonumber\\\fl&-&\frac{\dP^3\ddf^2\delta_{ij}\p^{i}\df\p^{j}\df}{a^2}+\frac{2\dP^4\alo\ddf\delta_{ij}\p^{i}\df\p^{j}\tho}{a^2}-\frac{\dP^5\alo^2\delta_{ij}\p^{i}\df\p^{j}\tho}{a^2}\Bigg)\nonumber\\\fl&+&P_{X\f}\Bigg(-\alo\df\ddf^2+2\dP\alo^2\df\ddf-2\dP\alt\df\ddf-\dP^2\alo^3\df+2\dP^2\alo\alt\df\nonumber\\\fl&-&\frac{2\dP\df\delta_{ij}\p^{i}\df\b^{j}}{a^2}-\frac{\alo\df\delta_{ij}\p^{i}\df\p^{j}\df}{a^2}+\frac{\df\ga_{ij}\p^{i}\df\p^{j}\df}{a^2}\nonumber\\\fl&-&\frac{2\df\ddf\delta_{ij}\p^{i}\df\p^{j}\tho}{a^2}+\frac{2\dP\alo\df\delta_{ij}\p^{i}\df\p^{j}\tho}{a^2}+\frac{2\dP\df\ga_{ij}\p^{i}\df\p^{j}\tho}{a^2}\nonumber\\\fl&-&\frac{2\dP\df\delta_{ij}\p^{i}\df\p^{j}\tht}{a^2}
\Bigg)-\frac{4H\alo\ga_{ij}\p^{i}\p^{j}\tho}{a^2}+\frac{4H\alt\ga_{ij}\p^{i}\p^{j}\tho}{a^2}\nonumber\\\fl&-&\frac{2H\alo\ga^{k}_{i}\ga_{jk}\p^{i}\p^{j}\tho}{a^2}-\frac{\alo^2\dot{\ga}_{ij}\p^{i}\p^{j}\tho}{a^2}+\frac{\alt\dot{\ga}_{ij}\p^{i}\p^{j}\tho}{a^2}-\frac{\alo\ga^{k}_{i}\dot{\ga}_{jk}\p^{i}\p^{j}\tho}{a^2}\nonumber\\\fl&+&\frac{\alo^2\p_{i}\p_{j}\tho\p^{i}\p^{j}\tho}{a^4}-\frac{\alt\p_{i}\p_{j}\tho\p^{i}\p^{j}\tho}{a^4}-\frac{2\alo\p_{i}\p_{j}\tho\p^{i}\p^{j}\tho}{a^4}\nonumber\\\fl&+&\frac{4H\alo\ga_{ij}\p^{i}\p^{j}\tht}{a^2}+\frac{2H\ga_{i}^{k}\ga_{jk}\p^{i}\p^{j}\tht}{a^2}+\frac{\alo\dot{\ga}_{ij}\p^{i}\p^{j}\tht}{a^2}+\frac{\ga_{i}^{k}\dot{\ga}_{jk}\p^{i}\p^{j}\tht}{a^2}\nonumber\\\fl&+&\frac{\p_{i}\p_{j}\tht\p^{i}\p^{j}\tht}{a^4}+\frac{2H\b^{i}\ga^{jk}\p_{k}\ga_{ij}}{a^2}+\frac{\b^{i}\dot{\ga}^{jk}\p_{k}\ga_{ij}}{a^2}-\frac{2H\alo\ga^{jk}\p^{i}\tho\p_{k}\ga_{ij}}{a^2}\nonumber\\\fl&-&\frac{\alo\dot{\gamma}^{jk}\p^{i}\tho\p_{k}\ga_{ij}}{a^2}+\frac{2H\ga^{jk}\p^{i}\tht\p_{k}\ga_{ij}}{a^2}+\frac{\dot{\ga}^{jk}\p^{i}\tht\p_{k}\ga_{ij}}{a^2}-\frac{\ga^{jk}\dot{\ga}^{l}_{j}\p^{i}\tho\p_{k}\ga_{jl}}{2a^2}\nonumber\\\fl&+&\frac{2\ga_{ij}\p^{i}\b^{j}\p^{2}\tho}{a^4}+\frac{2\ga_{ij}\p^{i}\p^{j}\tho\p^{2}\tht}{a^4}-\frac{\p_{i}\ga_{jk}\p^{i}\b^{j}\p^{k}\tho}{a^4}-\frac{\p_{j}\ga_{ik}\p^{i}\b^{j}\p^{k}\tho}{a^4}\nonumber\\\fl&+&\frac{\p^{j}\b^{i}\p_{k}\ga_{ij}\p^{k}\tho}{a^4}-\frac{2\ga_{jk}\p^{j}\b^{i}\p^{k}\p_{i}\tho}{a^4}+\frac{2\alo\ga_{jk}\p^{j}\p^{i}\tho\p^{k}\p_{i}\tho}{a^4}\nonumber\\\fl&+&\frac{\ga_{j}^{l}\ga_{kl}\p^{i}\p^{j}\tho\p^{k}\p_{i}\tho}{a^4}-\frac{4\ga_{jk}\p^{j}\p^{i}\tho\p^{k}\p_{i}\tht}{a^4}-\frac{2\ga_{ik}\p^{j}\b^{i}\p^{k}\p_{j}\tho}{a^4}\nonumber\\\fl&+&\frac{\beta^{i}\p_{i}\ga_{jk}\p^{k}\p^{j}\tho}{a^4}-\frac{2\alo\ga_{jk}\p^{2}\tho\p^{k}\p^{j}\tho}{a^4}-\frac{\ga_{j}^{l}\ga_{kl}\p^{2}\tho\p^{k}\p^{j}\tho}{a^4}\nonumber\\\fl&+&\frac{\alo\p_{i}\ga_{jk}\p^{i}\tho\p^{j}\p^{k}\tho}{a^4}-\frac{\ga^{l}_{j}\p_{i}\ga_{kl}\p^{i}\tho\p^{j}\p^{k}\tho}{a^4}+\frac{\p_{i}\ga_{jk}\p^{i}\tht\p^{j}\p^{k}\tho}{a^4}\nonumber\\\fl&-&\frac{2\b^{i}\p_{j}\ga_{ik}\p^{j}\p^{k}\tho}{a^4}+\frac{2\alo\p^{i}\tho\p_{j}\ga_{ik}\p^{k}\p^{j}\tho}{a^4}-\frac{2\p^{i}\tht\p_{j}\ga_{ik}\p^{j}\p^{k}\tho}{a^4}\nonumber\\\fl&+&\frac{\ga_{k}^{l}\p^{i}\tho\p_{j}\ga_{il}\p^{k}\p^{j}\tho}{a^{4}}+\frac{2\ga_{jk}\p^{2}\tho\p^{j}\p^{k}\tht}{a^4}+\frac{\p_{i}\ga_{jk}\p^{i}\tho\p^{j}\p^{k}\tht}{a^4}\nonumber\\\fl&-&\frac{2\p^{i}\tho\p_{j}\ga_{ik}\p^{j}\p^{k}\tho}{a^4}-\frac{2\ga^{kl}\p^{i}\tho\p^{2}\tho\p_{l}\ga_{ik}}{a^4}+\frac{2\ga^{j}_{l}\p^{i}\tho\p^{j}\p^{k}\tho\p_{l}\ga_{ik}}{a^4}\nonumber\\\fl&-&\frac{2H\ga^{j}_{i}\ga^{kl}\p^{i}\tho\p_{l}\ga_{jk}}{a^2}-\frac{\ga^{j}_{i}\dot{\ga}^{kl}\p^{i}\tho\p_{l}\ga_{jk}}{2a^2}+\frac{\ga^{l}_{i}\p^{i}\tho\p^{j}\p^{k}\tho\p_{l}\ga_{jk}}{a^4}\nonumber\\\fl&+&\frac{\delta^{kl}\p^{i}\tho\p^{j}\tho\p_{k}\ga_{i}^{t}\p_{l}\ga_{jt}}{2a^4}+\frac{\p^{i}\tho\p_{j}\ga_{kl}\p^{j}\tho\p^{l}\ga_{i}^{k}}{a^4}+\frac{\p^{i}\tho\p^{j}\tho\p_{k}\ga_{jl}\p^{l}\ga_{i}^{k}}{2a^4}\nonumber\\\fl&+&P_{X}\Bigg(\alo^2\ddf^2-\alt\ddf^2-2\dP\alo^3\ddf+4\dP\alo\alt\ddf+\dP^2\alo^4-3\dP^2\alo^2\alt\nonumber\\\fl&+&\dP^2\alt^2+\frac{2\df\delta_{ij}\b^{j}\p^{j}\df}{a^2}+\frac{2\dP\alo\delta_{ij}\b^{i}\p^{j}\df}{a^2}+\frac{2\dP\ga_{ij}\b^{i}\p^{j}\df}{a^2}\nonumber\\\fl&-&\frac{\alt\delta_{ij}\p^{i}\df\p^{j}\df}{a^2}+\frac{\alo\ga_{ij}\p^{i}\df\p^{j}\df}{a^2}-\frac{\delta^{kl}\ga{ik}\ga_{jl}\p^{i}\df\p^{j}\df}{2a^2}\nonumber\\\fl&+&\frac{2\alo\ddf\delta_{ij}\p^{i}\df\p^{j}\df}{a^2}+\frac{2\ddf\ga_{ij}\p^{i}\df\p^{j}\tho}{a^2}-\frac{2\dP\alo^2\delta_{ij}\p^{i}\df\p^{j}\tho}{a^2}\nonumber\\\fl&+&\frac{2\dP\alt\delta_{ij}\p^{i}\df\p^{j}\tho}{a^2}-\frac{2\dP\alo\ga_{ij}\p^{i}\df\p^{j}\tho}{a^2}-\frac{\dP\delta^{kl}\ga_{ik}\ga_{jl}\p^{i}\df\p^{j}\tho}{a^2}\nonumber\\\fl&-&\frac{2\ddf\delta_{ij}\p^{i}\df\p^{j}\tht}{a^2}+\frac{2\dP\alo\delta_{ij}\p^{i}\df\p^{j}\tht}{a^2}+\frac{2\dP\ga_{ij}\p^{i}\p^{j}\tht}{a^2}\nonumber\\\fl&+&\frac{\delta_{ik}\delta_{jl}\p^{i}\df\p^{j}\df\p^{k}\tho\p^{l}\tho}{a^4}\Bigg)+P_{XX}\Bigg(\frac{\ddf^4}{4}-3\dP\alo\ddf^3+\frac{17\dP^2\alo^2\ddf^2}{2}\nonumber\\\fl&-&4\dP^2\alt\ddf^2-9\dP^3\alo^3\ddf+10\dP^3\alo\alt\ddf+\frac{13\dP^4\alo^4}{4}-6\dP^4\alo^2\alt+\dP^4\alt^2\nonumber\\\fl&-&\frac{\dP^2\ddf\delta_{ij}\b^{i}\p^{j}\df}{a^2}+\frac{2\dP^3\alo\delta_{ij}\b^{i}\p^{j}\df}{a^2}-\frac{\ddf^2\delta_{ij}\p^{i}\df\p^{j}\df}{2a^2}\nonumber\\\fl&+&\frac{\dP\alo\delta_{ij}\p^{i}\df\p^{j}\df}{a^2}+\frac{\dP\ddf\delta_{ij}\p^{i}\df\p^{j}\df}{a^2}-\frac{\dP^2\alo^2\delta_{ij}\p^{i}\df\p^{j}\df}{2a^2}\nonumber\\\fl&+&\frac{\dP^2\alt\delta_{ij}\p^{i}\df\p^{j}\df}{a^2}-\frac{\dP^2\alo\ga_{ij}\p^{i}\df\p^{j}\df}{a^2}-\frac{3\dP\ddf\delta_{ij}\p^{i}\df\p^{j}\tho}{a^2}\nonumber\\\fl&+&\frac{8\dP^2\alo\ddf\delta_{ij}\p^{i}\df\p^{j}\tho}{a^2}+\frac{2\dP^2\ddf\ga_{ij}\delta_{ij}\p^{i}\df\p^{j}\tho}{a^2}+\frac{5\dP^3\alo^2\delta_{ij}\p^{i}\df\p^{j}\tho}{a^2}\nonumber\\\fl&+&\frac{2\dP^3\alt\delta_{ij}\p^{i}\df\p^{j}\tho}{a^2}-\frac{2\alo\ga_{ij}\p^{i}\df\p^{j}\tho\dP^3}{a^2}-\frac{2\dP^2\ddf\delta_{ij}\p^{i}\df\p^{j}\tht}{a^2}\nonumber\\\fl&+&\frac{2\dP^{3}\alo\delta_{ij}\p^{i}\df\p^{j}\tht}{a^2}+\frac{\left(\p^{2}\df\right)^2}{4a^4}+\frac{\dP\p^{2}\df\delta_{ij}\p^{i}\df\p^{j}\tht}{a^4}\nonumber\\\fl&+&\frac{\dP^2\delta_{ik}\delta_{jl}\p^{i}\df\p^{j}\df\p^{k}\tho\p^{l}\tho}{a^4}\Bigg)+\frac{\ga_{ik}\ga_{jl}\p^{i}\p^{j}\tho\p^{k}\p^{l}\tho}{a^{4}}-\frac{\ga_{ij}\ga_{kl}\p^{i}\p^{j}\tho\p^{k}\p^{l}\tho}{a^{4}}
\Bigg]
\eea

%%%%%%%%%%%%%%%%%%%%%%%%%%%%%%%%%%%%%%%%%%%%%%%%%%%%%%%%%%%%%%%%%%%%%%%%%%%%%%%%%%%%%%%%%%%%%%%%%%%%%%%%%%%%%%%%%%%%%
%%%%%%%%%%%%%%%%%%%%%%%%%%%%%%%%%%%%%%%%%%%%%%%%%%%%%%%%%%%%%%%%%%%%%%%%%%%%%%%%%%%%%%%%%%%%%%%%%%%%%%%%%%%%%%%%%%%%%%%%%%%%%%%%%%%%%%%%%%%%%%%%%%%%%%%%%%%%%%%%%%%%%%%%%%%%%%%%%%%%%%%%%%%%%%%%%%%%%%%%%%%%%%%%%%%%%%%%%%%%%%%%%%%%%%%%%%%%%%%%%%%%%%%%%%%%%%%%%%%%%%%%%%%%%%%%%%%%%%%%%%%%%%%%%%%%%%%%%%%%%%%%%%%%%%%%%%%%%%%%%%%%%%%%%%%%%%%%%%%%%%%%%%%%%%%%%%%%%%%%%%%%%%%%%%%%%%%%%%%%%%%%%%%%%%%%%%%%%%%%%%%%%%%%%%%%%%%%%%%%%%%%%%%%%%%%%%%%%%%%%%%%%%%%%%%%%%%%%%%%%%%%%%%%%%%%%%%%%%%%%%%%%%%%%%%%%%%%%%%%%%%%%%%%%%%%%%%%%%%%%%%%%%%%%%%%%%%%%%%%%%%%%%%%%%%%%%%%%%%%%%%%%%%%%%%%%%%%%%%%%%%%%%%%%%%%%%%%%%%%%%%%%%%%%%%%%%%%%%%%%%%%%%%%%%%%%%%%%%%%%%%%%%%%%%%%%%%%%%%%%%%%%%%%%%%%%%%%%%%%%%%%%%%%%%%%%%%%%%%%%%%%%%%%%%%%%%%%%%%%%%%%%%%%%%%%%%%%%%%%%%%%%%%%%%%%%%%%%%%%%%%%%%%%%%%%%%%%%%%%%%%%%%%%%%%%%%%%%%%%%%%%%%%%%%%%%%%%%%%%%%%%%%%%%%%%%%%%%%%%%%%%%%%%%%%%%%%%%%%%%%%%%%%%%%%%%%%%%%%%%%%%%%%%%%%%%%%%%%%%%%%%%%%%%%%%%%%%%%%%%%%%%%%%%%%%%%%%%%%%%%%%%%%%%%%%%%%%%%%%%%%%%%%%%%%%%%%%%%%%%%%%%%%%%%%%%%%%%%%%%%%%%%%%%%%%%%%%%%%%%%%%%%%%%%%%%%%%%%%%%%%%%%%%%%%%%

\setcounter{equation}{0}
\def\theequation{C\arabic{equation}}
\section*{Appendix C. Details about the computation of two-vertex diagrams}\label{2ver}

We will provide a summary of the main steps of how to compute the two-vertex diagrams. \\
We first expand the IN-IN formula as in Eq.~(\ref{eq3})
\bea \label{36}
\fl
\langle\df_{\vec{k_{1}}}\df_{\vec{k_{2}}}\rangle_{*}&\supset&\frac{(-i)^{2}}{2}\Big\langle T\Big[\df_{\vec{k_{1}}}\df_{\vec{k_{2}}}\int^{\eta^{*}}_{- \infty} d^3x^{'} d \epr a^{4}(\epr)\Big(H_{int}^{+}(\epr)-H_{int}^{-}(\epr)\Big)\nonumber \\\fl
&\times& \int^{\eta^{*}}_{- \infty} d^3x^{''} d \eps a^{4}(\eps)\left(H_{int}^{+}(\eps)-H_{int}^{-}(\eps)\right)\Big]\Big\rangle\nonumber\\\fl
&=& \frac{(-i)^{2}}{2}\Big\langle T\Big[\df_{\vec{k_{1}}}\df_{\vec{k_{2}}}\Big(A+B+C+D\Big)\Big]\Big\rangle,
\eea 
where
\bea
A \equiv \int^{\eta^{*}}_{- \infty} d^3x^{'} d \epr a^{4}H_{int}^{+}\int^{\eta^{*}}_{- \infty} d^3x^{''} d \eps a^{4}H_{int}^{+},\\
B \equiv \int^{\eta^{*}}_{- \infty} d^3x^{'} a^{4}d \epr H_{int}^{-}
\int^{\eta^{*}}_{- \infty} d^3x^{''} d \eps a^{4}H_{int}^{-},\\
C \equiv -\int^{\eta^{*}}_{- \infty} d^3x^{'} d \epr a^{4}H_{int}^{+}\int^{\eta^{*}}_{- \infty} d^3x^{''} d \eps a^{4}H_{int}^{-},\\
D \equiv -\int^{\eta^{*}}_{- \infty} d^3x^{'} d \epr a^{4}H_{int}^{-}
\int^{\eta^{*}}_{- \infty} d^3x ^{''}d \eps a^{4}H_{int}^{+}.
\eea
The plus and minus signs arise from having both time-ordering and anti time-ordering operators and give rise to different Feynman contraction rules for the field operators. After expanding the field operators in $H_{int}^{(3)}$ in their Fourier modes, contractions can be made among creation and annihilation operators. We can then integrate over space ($d^3 x$) using the $\delta$ function properties and, for each diagram, we are eventually left with an integral over the internal momentum running in the loop combined with a double time-integration. Before integrating in time, it can be convenient to perform a change of variables $x\equiv -k \eta$. As to the momentum integrals, it can be helpful to consider the following scheme that takes into account the momentum dependence of both the integrand function and the external legs as well as every momentum factor expected in the final expression:

\begin{itemize}

\item every eigenfunction ($\df_{q}$ or $\ga_{q}$) generates a $(2q)^{-3/2}$ factor (see Eqs.~(\ref{uk})-(\ref{gk}));  

\item spatial derivatives $\p_{i}$ and $\p^{-2}$ respectively produce a $q_{i}$ and a $q^{-2}$ factors. No $i$ factors or minus signs need to be included since the spatial derivatives always appear in an even number in our vertices;

\item temporal derivatives $\ddf_{q}$ or $\dot{\ga}_{q}$ originates a $-(q/k)^2$ factor, e.g.:
\bea\fl
\ddf_{q}&=&\frac{1}{a}\df^{'}\simeq-H\eta\times\frac{H}{(2q)^{3/2}}e^{-i q c_{s}\eta}q^2 c_{s}^2 \eta\nonumber\\\fl&=&-H\eta\times\frac{H}{(2q)^{3/2}}e^{-i\frac{q}{k}(k c_{s}\eta)}\left(\frac{q}{k}\right)^2 k^2 c_{s}^2 \eta\nonumber\\\fl&=& -\frac{H^2}{(2q)^{3/2}}e^{i\frac{q}{k}x}\left(\frac{q}{k}\right)^2 x^2
\eea
where $x\equiv -k c_{s}\eta$;
\item every vertex present in a given diagram provides an extra $k^3$ factor
\bea\fl
\int d^{4}x\sqrt{-g}\rightarrow\int d\eta a^{4}(\eta)=\int \frac{d\eta (-k c_{s})}{H^4(k c_{s}\eta)^4}(-k c_{s})^3=-\int \frac{dx}{H^4x^4}(k c_{s})^3 ;
\eea
\item all $a^{n}$ factors present in the verteces $V_{i}$ contribute with extra $k^{n}$ factor;
\item the polarization tensors are also responsible for producing a momentum dependence: they combine with one another according to the rules listes in Appendix C, Eqs.(\ref{pol1}) through (\ref{pol2}). 
\end{itemize}

\noindent The final results for the leading two-vertex diagrams with scalar loops are
\bea\label{fr1}\fl
G_{aa}(x^*)F_{aa}(k)&=&\frac{1}{60 k^3}\left(-4-4x^{*2}-25x^{*4}+x^{*6}\right)\ln\left(\Lambda/H_{*}\right),\\\label{fr2}\fl  
G_{ab}(x^*)F_{ab}(k)&=&\frac{1}{120 k^3}\left(19-71x^{*2}+41x^{*4}\right)\ln\left(\Lambda/H_{*}\right),\\\label{fr3}\fl
G_{bb}(x^*)F_{bb}(k)&=& \frac{1}{120 k^3}\left(32+384x^{*2}+87x^{*4}+4x^{*6}\right)\ln\left(\Lambda/H_{*}\right),
\eea
where the functions $G$ and $F$ were introduced in Eq.~(\ref{spiego0}).\\

%%%%%%%%%%%%%%%%%%%%%%%%%%%%%%%%%%%%%%%%%%%%%%%%%%%%%%%%%%%%%%%%%%%%%%%%%%%%%%%%%%%%%%%%%%%%%%%%%%%%%%%%%%%%%%%%%%%%%
%%%%%%%%%%%%%%%%%%%%%%%%%%%%%%%%%%%%%%%%%%%%%%%%%%%%%%%%%%%%%%%%%%%%%%%%%%%%%%%%%%%%%%%%%%%%%%%%%%%%%%%%%%%%%%%%%%%%%
%%%%%%%%%%%%%%%%%%%%%%%%%%%%%%%%%%%%%%%%%%%%%%%%%%%%%%%%%%%%%%%%%%%%%%%%%%%%%%%%%%%%%%%%%%%%%%%%%%%%%%%%%%%%%%%%%%%%%
%%%%%%%%%%%%%%%%%%%%%%%%%%%%%%%%%%%%%%%%%%%%%%%%%%%%%%%%%%%%%%%%%%%%%%%%%%%%%%%%%%%%%%%%%%%%%%%%%%%%%%%%%%%%%%%%%%%%%
%%%%%%%%%%%%%%%%%%%%%%%%%%%%%%%%%%%%%%%%%%%%%%%%%%%%%%%%%%%%%%%%%%%%%%%%%%%%%%%%%%%%%%%%%%%%%%%%%%%%%%%%%%%%%%%%%%%%

\setcounter{equation}{0}
\def\theequation{D\arabic{equation}}
\section*{Appendix D. Details about the computation of one-vertex diagrams}\label{1ver}

Let us now consider in details the contributions and analytic results from one-vertex diagrams. We will first recall the remark made at the end of Sec.~4 about the one-vertex diagrams with tensor loops. This is important in order to quickly recognize the diagrams that provide contributions that can be absorbed in a renormalization constant and whose computation can be therefore avoided. Indeed, the loop contributions that are interesting are those that originate from a momentum loop integral with a real dependence from the external momentum, i.e. that result in a logarithmic function of momentum in the final result. 
%, which can only occur if the vertex includes some inverse derivative applied to a product of wavefunctions, such as for instance $\p^{-2}\left[\ddf^2\right]$ or $\p^{-2}\left[\left(\p_{j}\df\right)\ddf\right]$. 
On the other hand, a four-leg vertex that is responsible for a diagram with the following mathematical form
\be\label{pol}
Pol(x^{*})\int_{0}^{+\infty} dq q^{\alpha}
\ee
(where $Pol(x^{*})$ is a polynomial function of the horizon crossing time $x^{*}\equiv -k \eta^{*}$, $\alpha$ is a positive integer), produces a disconnected-like contribution to the two-point function which can be absorbed into a left-over renormalization constant. For this reason, tensor vertices have not been included in Eq.~(\ref{s4}) and have not been taken into account in our one-vertex diagrams computations: by looking at the complete expression for the Lagrangian provided in Appendix B (Eq.~(\ref{gr11})), it is easy to verify that all of the graviton non zero contributions have a form as in Eq.~(\ref{pol}). For the same reason, the only leading order scalar diagrams that we have taken into account as far as the one-loop contributions are concerned, are the ones from the vertices given in Eqs.~(\ref{vertice01}) through (\ref{vertice41}), which we rewrite below
\bea\label{vertice0}
V_{(0)}\equiv \frac{P_{XXX}\Sigma\fd^2}{H}\tilde{\alt}\ddf^2,\\\label{vertice1}
V_{(1)}\equiv 24\frac{\lambda \Sigma}{H\fd^4}\left(\p_{i}\p_{j}\tilde{\tht}\right)\left(\p_{i}\tilde{\beta}_{j}\right),\\\label{vertice2}
V_{(2)}\equiv -12\frac{\lambda\Sigma}{H\fd^4}\tilde{\alt}\ddf^2,\\\label{vertice3}
V_{(3)}\equiv -\frac{P_{XX}\Sigma}{a^2 H}\tilde{\alt}\left(\delta_{ij}\p_{i}\df\p_{j}\df\right),\\\label{vertice4}
V_{(4)}\equiv -6\frac{P_{XX}\lambda}{H}\ddf\delta_{ij}\left(\p_{i}\df\right)\left(\p_{j}\tilde{\tht}\right).
\eea
These can be rewritten as
\bea
%V_{0A}=\p^{-2}\left(\p^{2}\df\ddf\right)\ddf^2\\
%V_{0B}=\p^{-2}\left(\delta_{ij}\p_{i}\df\p_{j}\ddf\right)\ddf^2\\
V_{(1A)}&\simeq&\p_{i}\p_{j}\p^{-2}\left(\ddf\ddf\right)\p_{i}\p_{j}\p^{-4}\left(\p^{2}\df\ddf\right), \\
V_{(1B)}&\simeq&\p_{i}\p_{j}\p^{-2}\left(\ddf\ddf\right)\p_{i}\p_{j}\p^{-4}\left(\delta_{lm}\p_{l}\df\p_{m}\ddf\right), \\
V_{(1C)}&\simeq&\p_{i}\p_{j}\p^{-2}\left(\ddf\ddf\right)\p_{i}\p^{-2}\left(\p_{j}\df\ddf\right), \\
V_{(3A)}&\simeq&\p^{-2}\left(\p^{2}\df\ddf\right)\delta_{ij}\p_{i}\df\p_{j}\df, \\
V_{(3B)}&\simeq&\p^{-2}\left(\delta_{ij}\p_{i}\df\p_{j}\ddf\right)\delta_{lm}\p_{l}\df\p_{m}\df, \\
V_{(4)}&\simeq&\ddf\delta_{ij}\p_{i}\df\p_{j}\p^{-2}\left(\ddf\ddf\right).
\eea
As to $V_{(0)}$ and $V_{(2)}$, using the definition of $\lambda$ and $\Sigma$ from Eqs.(\ref{m}) and (\ref{n}) and referring back to Table~1, it is easy to verify that they cancel each other to leading order in the slow-variation parameters.\\
\noindent From the expansion of the Schwinger-Keldysh formula (\ref{eq3}) we have 
\be
\langle\df_{\vec{k_{1}}}\df_{\vec{k_{2}}}\rangle_{*}\supset i\int d\epr \left\langle\left[H_{int}^{(4)}(\epr),\df_{\vec{k_{1}}}\df_{\vec{k_{2}}}\right]\right\rangle.
\ee 
Before proceeding with the derivation of the analytic expressions of the diagrams corresponding to $V_{(1)}$, $V_{(3)}$ and $V_{(4)}$, it is important to notice that, for each one of them, both ``regular'' and ``singular'' contractions exists. To understand what we mean, let us consider the diagrams originating from $V_{(1A)}\simeq\p_{i}\p_{j}\p^{-2}\left(\ddf\ddf\right)\p_{i}\p_{j}\p^{-4}\left(\p^{2}\df\ddf\right)$. If we call $\vec{q}_{1}$, $\vec{q}_{2}$, $\vec{q}_{3}$ and $\vec{q}_{4}$ respectively the momenta associated with the four internal legs, we count three possible different types of contractions, i.e. 
\bea
\widehat{\vec{k}_{1}\vec{q}_{2}}\times\widehat{\vec{k}_{2}\vec{q}_{1}}\times\widehat{\vec{q}_{3}\vec{q}_{4}},\\
\widehat{\vec{k}_{1}\vec{q}_{1}}\times\widehat{\vec{k}_{2}\vec{q}_{3}}\times\widehat{\vec{q}_{2}\vec{q}_{4}},\\
\widehat{\vec{k}_{1}\vec{q}_{4}}\times\widehat{\vec{k}_{2}\vec{q}_{1}}\times\widehat{\vec{q}_{2}\vec{q}_{3}},
\eea
where the $ \hat{} $ symbol stands for a contraction between the two wavefunctions to which the momenta are associated. It is easy to check that the second and the third permutations respectively provide a factor
\bea
\frac{k_{2}^2\left[\left(\vec{q}-\vec{k}_{1}\right)\cdot\left(\vec{q}+\vec{k}_{2}\right)\right]^2}{|\vec{q}-\vec{k}_{1}|^2|\vec{q}+\vec{k}_{2}|^4},\\
\frac{q^2\left[\left(\vec{q}-\vec{k}_{2}\right)\cdot\left(\vec{q}+\vec{k}_{1}\right)\right]^2}{|\vec{q}-\vec{k}_{2}|^2|\vec{q}+\vec{k}_{1}|^4},
\eea
inside the momentum integral, where $\vec{q}$ is the momentum running inside the loop. The first permutation instead gives a factor
\bea
\frac{q^2\left[\left(\vec{k}_{1}+\vec{k}_{2}\right)\cdot\left(\vec{q}-\vec{q}\right)\right]^2}{|\vec{k}_{1}+\vec{k}_{2}|^2|\vec{q}-\vec{q}|^4},
\eea
which is apparently singular, also considering momentum conservation.\\
We will now show how these ``singular'' diagrams end up not contributing to the final result at all.\\

\begin{figure}\centering
 \includegraphics[width=0.4\textwidth]{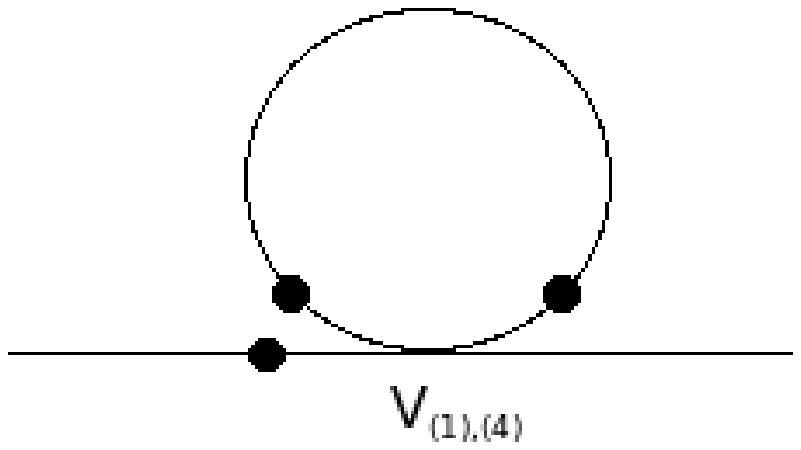}\vspace{0.05\textwidth}
\hspace{0.1\textwidth}
 \includegraphics[width=0.4\textwidth]{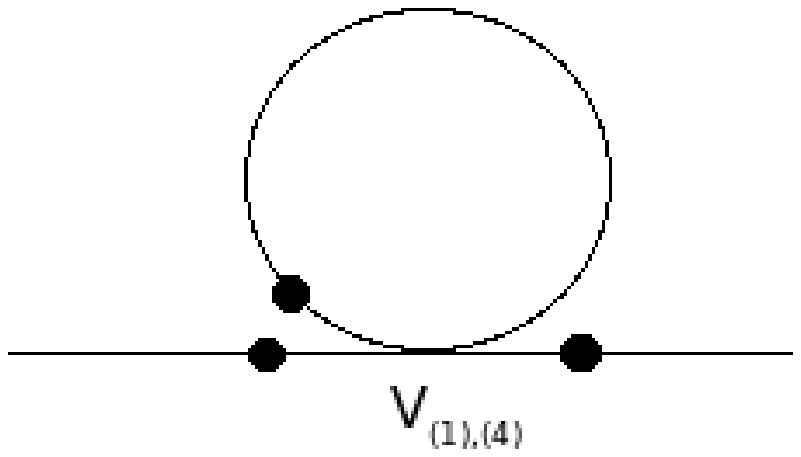}
\vspace{0.05\textwidth}
 \includegraphics[width=0.4\textwidth]{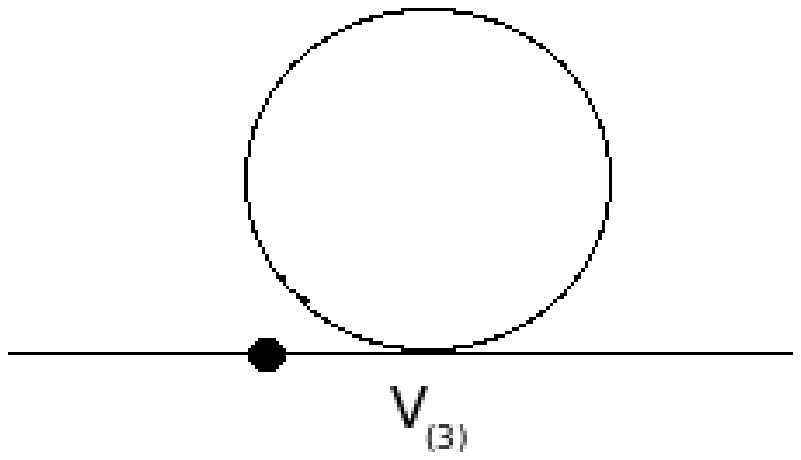}
\hspace{0.1\textwidth}
 \includegraphics[width=0.4\textwidth]{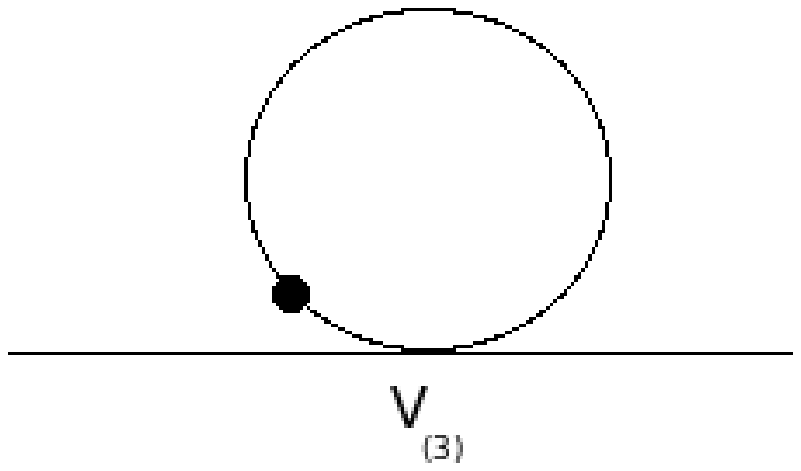}
\caption{ \label{Fig1} Diagrammatic representation of the one-vertex interaction diagrams. The dots on each end of a line indicate temporal derivatives of the corresponding wavefunctions.}
\end{figure}

\noindent From $V_{1A}$ we get a singular contribution
\bea\fl
\int d^{3}q \Big[2\int dx T_{b}(x)\frac{\left[(\vec{k_{1}}+\vec{k_{2}})\cdot(\vec{q}-\vec{q})\right]^{2}k^4 q^4}{{|\vec{k_{1}}+\vec{k_{2}}|}^{2}{|\vec{q}-\vec{q}|}^{4}}+2\int dx T_{a}(x)\frac{\left[(\vec{k_{1}}+\vec{k_{2}})\cdot(\vec{q}-\vec{q})\right]^{2}k^4 q^4}{{|\vec{k_{1}}+\vec{k_{2}}|}^{4}{|\vec{q}-\vec{q}|}^{2}}\Big]\nonumber\\
\eea
%(where $T_{a}\equiv \frac{x^2}{k^3}e^{2i(x-x^{*})}(1-ix^{*})^2(1+ix)$ and $T_{b}\equiv\frac{x^2}{k^3}e^{2i(x-x^{*})}(1-ix^{*})^2(1+i\frac{q}{k}x)$ (\textbf{da controllare})) 
(where we have condensed the products of wavefunctions inside the symbols $T_{a}$ and $T_{b}$) which is equal and opposite to the one from $V_{1B}$. Similarly, we verified that the singular diagrams from $V_{3A}$
\bea\
\int d^{3}q \left[-2\int dx T_{c}(x)\frac{(k_{1}^{2}k_{2}^{2})q^2}{{|\vec{k_{1}}+\vec{k_{2}}|}^{2}}+2\int dx T_{d}(x)\frac{q^4(\vec{k_{1}}\cdot\vec{k_{2}})}{{|\vec{q}-\vec{q}|}^{2}}\right]
\eea
%(where $T_{c}(x)\equiv \frac{1}{k}e^{2i(x-x^{*})}(1-ix^{*})^2(1+ix)^2(1+i\frac{q}{k}x)$ and $T_{d}(x)\equiv \frac{1}{k}e^{2i(x-x^{*})}(1-ix^{*})^2(1+ix)|1-i\frac{q}{k}x|^2\\T_{e}(x)\equiv \frac{x^2}{k^3}e^{2i(x-x^{*})}(1-ix^{*})^2(1-i\frac{q}{k}x)$ (\textbf{da controllare})) 
cancels the ones from $V_{3B}$. Finally we can show that the singular contributions from $V_{1C}$ and $V_{4}$ (surviving after we sum up over the permutations) 
\bea
V_{1C}&\rightarrow& 
\int d^{3}q \left[-2\frac{\left[(\vec{k_{1}}+\vec{k_{2}})\cdot(\vec{q}-\vec{q})\right]\left[\vec{q}\cdot(\vec{k_{1}}+\vec{k_{2}})\right]k^4 q^2}{{|\vec{k_{1}}+\vec{k_{2}}|}^{2}{|\vec{q}-\vec{q}|}^{2}}\right]
\\
V_{4} &\rightarrow&\int d^{3}q \left[\frac{k^4 q^2 \vec{q}\cdot(\vec{k_{1}}+\vec{k_{2}})}{{|\vec{k_{1}}+\vec{k_{2}}|}^{2}}\right]
\eea
give a zero contribution as well. As to $V_{4}$, one can write it as
\bea\label{symm}
\frac{k^4}{\epsilon^2}\vec{\epsilon}\cdot\left[\int d^{3}q q^2 \vec{q}\right]
\eea
where we define $\vec{\epsilon}\equiv\vec{k}_{1}+\vec{k}_{2}$. From Eq.(\ref{symm}) it is apparent that the integral goes to zero for symmetry reasons. The same argument applies to $V_{1C}$. Let us write it as
\bea
\frac{\vec{\epsilon}\cdot\vec{\delta}}{\epsilon^2 \delta^2}\vec{\epsilon}\cdot\left[\int d^3 q \vec{q}\right]
\eea 
where $\vec{\delta}\equiv\vec{q}-\vec{q}$ and we assumed that $\delta$ is independent of the integration variable q and so can be taken outside the integral.\\
We are now left with dealing with the ``regular'' diagrams. The time integrals are diagrammatically represented in Fig.~(\ref{Fig1}). The momentum integrals can be easily performed. We report the final results 
\bea\fl
G_{1}(x^{*})F_{1}(k)&=&\frac{1}{30}\left(-5-5 {x}^{*2}-18 {x}^{*4}\right) \frac{\ln\left(\Lambda/H_{*}\right)}{k^3},\\\fl
G_{3}(x^{*})F_{3}(k)&=& \frac{1}{120} \left(-405-745 x^{*2}-366 x^{*4}\right) \frac{\ln\left(\Lambda/H_{*}\right)}{k^3} ,\\\fl
G_{4}(x^{*})F_{4}(k)&=& \frac{1}{60} \left(5+5 {x}^{*2}+18 {x}^{*4}\right) \frac{\ln\left(\Lambda/H_{*}\right)}{k^3}.
\eea
%\textbf{credo da moltiplicare ancora per i loro fattori esterni, ad esempio 1 e 4 vanno moltiplicati per 16 e -16 rispettivamente, 3 va moltiplicato per 4}. 
The $\ln\left(\Lambda/H_{*}\right)$ terms come from ultraviolet divergences.

%%%%%%%%%%%%%%%%%%%%%%%%%%%%%%%%%%%%%%%%%%%%%%%%%%%%%%%%%%%%%%%%%%%%%%%%%%%%%%%%%%%%%%%%%%%%%%%%%%%%%%%%%%%%%%%%%%%%%%
%%%%%%%%%%%%%%%%%%%%%%%%%%%%%%%%%%%%%%%%%%%%%%%%%%%%%%%%%%%%%%%%%%%%%%%%%%%%%%%%%%%%%%%%%%%%%%%%%%%%%%%%%%%%%%%%%%%%%%
%%%%%%%%%%%%%%%%%%%%%%%%%%%%%%%%%%%%%%%%%%%%%%%%%%%%%%%%%%%%%%%%%%%%%%%%%%%%%%%%%%%%%%%%%%%%%%%%%%%%%%%%%%%%%%%%%%%%%%
%%%%%%%%%%%%%%%%%%%%%%%%%%%%%%%%%%%%%%%%%%%%%%%%%%%%%%%%%%%%%%%%%%%%%%%%%%%%%%%%%%%%%%%%%%%%%%%%%%%%%%%%%%%%%%%%%%%%%%
%%%%%%%%%%%%%%%%%%%%%%%%%%%%%%%%%%%%%%%%%%%%%%%%%%%%%%%%%%%%%%%%%%%%%%%%%%%%%%%%%%%%%%%%%%%%%%%%%%%%%%%%%%%%%%%%%%%%%
%%%%%%%%%%%%%%%%

\setcounter{equation}{0}
\def\theequation{E\arabic{equation}}
\section*{Appendix E. Polarization tensor equations}\label{polarization}

The following relations can be useful in the calculations involving the tensor modes
\bea\label{pol1}
\ep_{ij}^{\lambda}(\hat{q})k^{i}k^{j}=\frac{k^2 \sin^2\theta}{\sqrt{2}},\\
k^{i}k^{j}\ep_{ik}^{\lambda}(\hat{q})\ep_{kj}^{*\lambda}(\hat{q})=2k^2\sin^2\theta,\\
\ep_{ij}^{\lambda}(\hat{z})\ep_{ij}^{\lambda}(\hat{q})=\frac{{\left(k^2-(q-z)^2\right)}^{2}}{4q^2 z^2},\\\label{pol2}
\ep_{ij}^{\lambda}(\hat{z})\ep_{jl}^{\lambda}(\hat{q})k^{i}k^{l}=\ep_{ij}^{\lambda}(\hat{z})\ep_{jl}^{\lambda}(\hat{q})q^{i}k^{l}=-\frac{\left(k^2-(q-z)^2\right)^{2}\left(-k^2+(q+z)^2\right)}{8q^2 z^2},\nonumber\\
\eea
where $z\equiv |\vec{q}-\vec{k}|$ and sums are taken over repeated (spatial and polarization) indices. These equations can be easily derived as follows. We can choose a spatial coordinate frame so that the vector $\vec{k}$ points along the third spatial direction, then the components of $\ep_{ij}(\hat{k})$ are $\ep_{11}=-\ep_{22}=1/\sqrt{2}$, $\ep_{12}=\ep_{21}=\pm i/\sqrt{2}$, $\ep_{13}=\ep_{31}=0$ (the plus and minus signs refer to the two polarization states of the tensor). The tensor $\ep_{ij}(\hat{q})$ is obtained by rotation along the direction of $\hat{q}=(\sin\theta \cos\phi,\sin\theta\sin\phi,\cos\theta)$. The result is 
\bea
\ep_{11}=\frac{\cos^2\theta}{\sqrt{2}},\,\,\,\,\ep_{22}=-\frac{1}{\sqrt{2}},\,\,\,\,\ep_{33}=\frac{\sin^2\theta}{\sqrt{2}},\,\,\,\,\,\nonumber\\
\ep_{12}=\ep_{21}=\pm\frac{ i \cos\theta}{\sqrt{2}},\,\,\,\,\,
\ep_{13}=\ep_{31}=\frac{-\sin\theta\cos\theta}{\sqrt{2}},\,\,\,\,\,\nonumber\\\ep_{23}=\ep_{32}=\mp \frac{i \sin\theta}{\sqrt{2}}
\eea
where we set $\phi=0$.\\
Similarly, $\epsilon_{ij}(\hat{z})$ is obtained by rotating $\epsilon_{ij}(\hat{z})$ along the direction of $\hat{z}=(\sin\theta_{z},0,\cos\theta_{z})$, where $\sin\theta_{z}\equiv \sin\theta (q/z)$ and $\cos\theta_{z}\equiv (q\cos\theta-k)/z$.\\
Using the matrices $\epsilon_{ij}(\hat{q})$ and $\epsilon_{ij}(\hat{z})$ thus constructed and considering the polarization tensor orthogonality condition $q^{i}\ep_{ij}(\hat{q})=0$, Eq.(\ref{pol1}) through (\ref{pol2}) can be straightforwadly derived.

%%%%%%%%%%%%%%%%%%%%%%%%%%%%%%%%%%%%%%%%%%%%%%%%%%%%%%%%%%%%%%%%%%%%%%%%%%%%%%%%%%%%%%%%%%%%%%%%%%%%%%%%%%%%%%%%%%%%%%%%%%%%%%%%%%%%%%%%%%%%%%%%%%%%%%%%%%%%%%%%%%%%%%%%%%%%%%%%%%%%%%%%%%%%%%%%%%%%%%%%%%%%%%%%%%%%%%%%%%%%%%%%%%%%%%%%%%%%%%%%%%%%%%%%%%%%%%%%%%%%%%%%%%%%%%%%%%%%%%%%%%%%%%%%%%%%%%%%%%%%%%%%%%%%%%%%%%%%%%%%%%%%%%%%%%%%%%%%%%%%%%%%%%%%%%%%%%%%%%%%%%%%%%%%%%%%%%%%%%%%%%%%%%%%%%%%%%%%%%%%%%%%%%%%%%%%%%%%%%%%%%%%%%%%%%%%%%%%%%%%%%%%%%%%%%%%%%%%%%%%%%%%%%%%%%%%%%%%%%%%%%%%%%%%%%%%%%%%%%%%%%%%%%%%%%%%%%%%%%%%%%%%%%%%%%%%%%%%%%%%%%%%%%%%%%%%%%%%%%%%%%%%%%%%%%%%%%%%%%%%%%%%%%%%%%%%%%%%%%%%%%%%%%%%%%%%%%%%%%%%%%%%%%%%%%%%%%%%%%%%%%%%%%%%%%%%%%%%%%%%%%%%%%%%%%%%%%%%%%%%%%%%%%%%%%%%%%%%%%%%%%%%%%%%%%%%%%%%%%%%%%%%%%%%%%%%%%%%%%%%%%%%%%%%%%%%%%%%%%%%%%%%%%%%%%%%%%%%%%%%%%%%%%%%%%%%%%%%%%%%%%%%%%%%%%%%%%%%%%%%%%%%%%%%%%%%%%%%%%%%%%%%%%%%%%%%%%%%%%%%%%%%%%%%%%%%%%%%%%%%%%%%%%%%%%%%%%%%%%%%%%%%%%%%%%%%%%%%%%%%%%%%%%%%%%%%%%%%%%%%%%%%%%%%%%%%%%%%%%%%%%%%%%%%%%%%%%%%%%%%%%%%%%%%%%%%%%%%%%%%%%%%%%%%%%%%%%%%%%%%%%%%%%%%%%%%%%%%%%%%%%%%%%%%%%%%%%%%%%%%%%%%%%%%%%%%%%%%%%%%%%%%%%%%%%%%%%%%%%%%%%%%%%%%%%%%%%%%%%%%%%%%%%%%%%%%%%%%%%%%%%%%%%%%%%%%%%%%%%%%%%%%%%%%%%%%%%%%%%%%%%%%%%%%%%%%%%%%%%%%%%%%%%%%%%%%%%%%%%%%%%%%%%%%%%%%%%%%%%%%%%%%%%%%%%%%%%%%%%%%%%%%%%%%%%%%%%%%%%%%%%%%%%%

\end{document}